\documentclass[useAMS,usenatbib]{mn2e}
\usepackage{graphicx}
\usepackage{aas_macros}

\newcommand{\kms}{\mbox{${\rmn{km}\,\rmn{s}}^{-1}$}}

\newcommand{\Msolar}{\mbox{${M}_{\sun}$}}
\newcommand{\Lsolar}{\mbox{${L}_{\sun}$}}
\newcommand{\Rsolar}{\mbox{${R}_{\sun}$}}

\newcommand{\logg}{\mbox{$\log\,{\rmn{g}}$}}
\newcommand{\Teff}{\mbox{T$_{\rmn{eff}}$}}

\title[EL CVn-type binaries -- Discovery of 17 systems]{EL CVn-type
binaries -- Discovery of 17 helium white dwarf precursors in bright
eclipsing binary star systems}
\author[P. F. L. Maxted et~al.]{P. F. L.~Maxted,$^{1}$\thanks{E-mail:
p.maxted@keele.ac.uk} 
S.~Bloemen,$^{2}$ U.~Heber,$^{3}$ S.~Geier,$^{3, 4}$ P.~J.~Wheatley,$^5$
\newauthor
T. R. Marsh,$^5$ 
E. Breedt,$^5$ D. Sebastian,$^6$ G. Faillace,$^7$ C. Owen,$^7$ 
D. Pulley,$^7$
\newauthor  
D. Smith,$^7$ U. Kolb,$^7$  C. A. Haswell,$^7$ J. Southworth$^{1}$, 
D.~R. Anderson,$^{1}$ \newauthor
B. Smalley,$^{1}$ A. Collier Cameron,$^8$ L. Hebb,$^9$ E.~K. Simpson,$^{10}$
R. G. West,$^5$
\newauthor
J. Bochinski,$^7$ R. Busuttil$^7$ and S. Hadigal$^{7, 11}$
  \\
$^1$Astrophysics Group,  Keele University, Keele, Staffordshire ST5 5BG, UK\\
$^2$Instituut voor Sterrenkunde, University of Leuven, Celestijnenlaan 200D,
3001, Heverlee, Belgium \\
$^3$Dr. Karl Remeis-Observatory \& ECAP, Sternwartstr. 7, 96049,
Bamberg, Germany \\
$^4$European Southern Observatory, Karl-Schwarzschild-Str. 2, 85748, Garching,
Germany\\
$^5$Department of Physics, University of Warwick, Coventry CV4 7AL, UK\\
$^6$Th\"{u}ringer Landessternwarte Tautenburg, Sternwarte 5, 07778, Tautenburg,
Germany\\
$^7$Department of Physical Sciences, The Open University, Walton Hall, Milton
Keynes MK7 6AA, UK\\
$^8$SUPA, School of Physics \& Astronomy, University of St Andrews, North
Haugh, KY16 9SS, St Andrews, Fife, Scotland, UK\\
$^9$Department of Physics and Astronomy, Vanderbilt University, Nashville, TN
37235, USA\\
$^{10}$Astrophysics Research Centre, School of Mathematics \& Physics, Queen's
University Belfast, Belfast BT7 1NN\\
$^{11}$International Space University, 1 rue Jean-Dominique Cassini, 67400
Illkirch-Graffenstaden, France
}

\date{To be inserted}
\pagerange{\pageref{firstpage}--\pageref{lastpage}} \pubyear{2009}

\begin{document}

\maketitle

\label{firstpage}

\begin{abstract}
 The star 1SWASP~J024743.37$-$251549.2  was recently discovered to be a
binary star in which an A-type dwarf star eclipses the
remnant of a disrupted red giant star (WASP\,0247$-$25\,B). The remnant is in
a rarely-observed state evolving to higher effective temperatures at nearly
constant luminosity prior to becoming a very low-mass white dwarf composed
almost entirely of helium, i.e., it is a pre-He-WD. We have used the WASP
photometric database to find 17 eclipsing binary stars with orbital periods
P=0.7\,--\,2.2\,d with similar lightcurves to 1SWASP~J024743.37$-$251549.2.
The only star in this group previously identified as a variable star is the
brightest one, EL~CVn, which we adopt as the prototype for this class of
eclipsing binary star. The characteristic lightcurves of EL~CVn-type stars
show  a total eclipse by an A-type dwarf star of a smaller, hotter star and a
secondary eclipse of comparable depth to the primary eclipse. We have used new
spectroscopic observations for 6 of these systems to confirm that the
companions to the A-type stars in these binaries have very low masses
($\approx 0.2\Msolar$). This includes the companion to EL~CVn which was not
previously known to be a pre-He-WD. EL~CVn-type binary star systems will
enable us to study the formation of very low-mass white dwarfs in great
detail, particularly in those cases where the pre-He-WD star shows non-radial
pulsations similar to those recently discovered in WASP0247$-$25\,B.
\end{abstract}


\section{Introduction}
 1SWASP~J024743.37$-$251549.2 (WASP\,0247$-$25 hereafter) is one of several
million bright stars ($8\la V \la 13$)  that have been observed by the Wide
Angle Search for Planets, (WASP, \citealt{2006PASP..118.1407P}).
\citet{2011MNRAS.418.1156M} showed that this eclipsing binary star contains an
A-type dwarf star (WASP\,0247$-$25\,A) and a pre-He-WD with a mass $\approx
0.2\Msolar$ (WASP\,0247$-$25\,B). The only other example of a similar
eclipsing binary star known at that time was the star  V209 in the globular
cluster $\omega$~Cen \citep{2007AJ....133.2457K}. This is an extremely unusual
object in which a pre-He-WD is eclipsed by a 0.945\Msolar\ star with an
effective temperature $\Teff=9370$\,K. The eclipsing binary star AW~UMa
may also contain a pre-He-WD, although the interpretation of this system is
complicated by an equatorial belt of material that makes the lightcurve of
this binary look like that of a W~UMa-type contact binary star
\citep{2008MNRAS.386..377P}. More recently, \citet{2012Natur.484...75P} have
discovered that the star OGLE-BLG-RRLYR-02792 is an eclipsing binary star with
an orbital period of 15.2 days in which a pre-He-WD star with a mass of
0.26\Msolar\ shows RR-Lyrae-type pulsations with a period of 0.63 days. Some
very young helium white dwarfs in eclipsing binary systems have been
identified using Kepler photometry \citep{2010ApJ...713L.150R,
2010ApJ...715...51V, 2011ApJ...728..139C, 2012ApJ...748..115B}. These stars
are difficult to study because they are much fainter than their A-type and
B-type companion stars. In contrast, HZ~22 \citep{1978A&A....70..451S} and
HD~188112 \citep{2003A&A...411L.477H} are also known to be pre-He-WD, but the
companions to these non-eclipsing binary stars are too faint to have been
detected so far, and so it has not been possible to derive precise,
model-independent masses for these stars. 

  Low-mass white dwarf stars ($M\la 0.35\Msolar$) are the product of binary
star evolution \citep{1993PASP..105.1373I, 1995MNRAS.275..828M}. Various
evolution channels exist, but they are generally the result of mass transfer
from an evolved main sequence star or red giant star onto a companion star.
Towards the end of the mass transfer phase the donor star will have a
degenerate helium core. This ``stripped red giant star'' does not have 
sufficient mass to undergo a helium flash, and so the white dwarf that emerges
has an anomalously low mass and is composed almost entirely of helium. For
this reason, they are known as helium white dwarfs (He-WD).  If the companion
to the red giant is a neutron star then the mass transfer is likely to be
stable so the binary can go on to become a low mass X-ray binary containing a
millisecond pulsar. Several millisecond radio pulsars are observed to have
low-mass white dwarf companions \citep{2008LRR....11....8L}. Many He-WD have
been identified in the Sloan Digital Sky Survey \citep{2007ApJ...660.1451K},
some with masses as low as 0.16\Msolar\ \citep{2012ApJ...751..141K}, and from
proper motion surveys \citep{2009A&A...506L..25K}.  Searches for radio pulsar
companions to these white dwarfs have so far found nothing, so the companion
stars are likely to also be  white dwarf stars \citep{2009ApJ...697..283A}.
This can be confirmed in those few cases where these binary white dwarf pairs
show eclipses \citep{2011ApJ...735L..30P, 2011ApJ...737L..23B}. Helium white
dwarfs (He-WD) can also be produced by mass transfer from a red giant onto a
main sequence star, either rapidly through unstable common-envelope evolution
or after a longer-lived ``Algol'' phase of stable mass transfer
\citep{1969A&A.....1..167R, 1970A&A.....6..309G, 2004A&A...419.1057W,
1993PASP..105.1373I, 2003MNRAS.341..662C, 2001ApJ...552..664N}. He-WDs may
also be the result of collisions in dense stellar environments such as the
cores of globular clusters \citep{2008ApJ...683.1006K}.

 The evolution of He-WDs is expected to be very different from more massive
white dwarfs.  If the timescale for mass loss from the red giant is longer
than the thermal timescale, then when mass transfer ends there will still be  
a thick layer of hydrogen surrounding the degenerate helium core. The mass of
the hydrogen layer depends on the total mass and composition of the star
\citep{2004ApJ...616.1124N}, but is typically 0.001\,--\,0.005\Msolar, much
greater than for typical white dwarfs (hydrogen layer mass $<10^{-4}$\Msolar).
The pre-He-WD then evolves at nearly constant luminosity towards higher
effective temperatures as a result of the gradual reduction in the hydrogen
layer mass through stable p-p chain shell burning. This pre-He-WD phase can
last several million years for lower mass stars with thicker hydrogen
envelopes. Towards the end of this phase p-p chain fusion becomes less
efficient and the star starts to fade and cool.

 The smooth transition from a pre-He-WD to a He-WD can be interrupted by one or
more phases of unstable CNO burning (shell flashes) for pre-He-WD with masses
$\approx 0.2$\,--\,0.3\Msolar\ \citep{1975MNRAS.171..555W,
1999A&A...350...89D}. These shell flashes substantially reduce the mass of
hydrogen  that remains on the surface.
 The mass range within which shell flashes are predicted to occur
depends on the assumed composition of the star and other details of the models
\citep{2001MNRAS.323..471A}.   The cooling timescale for He-WD that do not
undergo shell flashes is much longer than for those that do because their
thick hydrogen envelopes can support residual p-p chain fusion for several
Gyr. 

 \citet{2013Natur.498..463M} found  strong observational support for the assumption
that He-WD are born with thick hydrogen envelopes.  They found that only models
with thick hydrogen envelopes could simultaneously match their precise mass
and radius estimates for both WASP\,0247$-$25\,A and WASP\,0247$-$25\,B,
together with other observational constraints such as the orbital period and
the likely composition of the stars based on their kinematics. In addition,
they found that WASP\,0247$-$25\,B is a new type of variable star in which
a mixture of radial and non-radial pulsations produce multiple frequencies in
the lightcurve near 250 cycles/day. This opens up the prospect of using
asteroseismology to study the interior of this star, e.g., to measure its
internal rotation profile.

 The study by Maxted et al. of  WASP\,0247$-$25 clearly demonstrates that
finding a pre-He-WD in a bright eclipsing binary system makes it possible to
study these rarely-observed stars in great detail and so better understand all
low-mass white dwarfs. This will, in turn, improve our understanding of the
various exotic binary systems and extreme stellar environments in which
low-mass white dwarfs are found. Motivated by this discovery, we have used the
WASP photometric database to search for similar eclipsing binary systems to
WASP\,0247$-$25. Here we present an analysis of the WASP lightcurves and
other data for 17 new eclipsing binary stars, 16 of which are new discoveries,
and all of which are newly identified as binary systems containing a pre-He-WD.

\begin{table}
\caption{New EL CVn-type binary stars identified using the WASP archive.
Spectral types are taken from the SIMBAD database.}
\label{startable}
\begin{tabular}{@{{}}lrrl}
\hline
\multicolumn{1}{@{}l}{1SWASP} & \multicolumn{1}{l}{P[d]} &\multicolumn{1}{l}{V}& Notes \\    
\hline
J013129.76$+$280336.5 & 1.882 & 10.9 & TYC~1755-509-1   \\
J034623.68$-$215819.5 & 0.928 &  9.6 & HD~23692, A4\,IV \\
J035838.33$-$311638.3 & 2.189 & 11.4 & CD$-$31~1621, A3 \\
J084356.46$-$113327.5 & 0.793 & 10.7 & TYC~5450-1192-1  \\
J084558.78$+$530209.8 & 0.844 & 12.9 &                  \\
J093957.74$-$191941.2 & 1.073 & 12.1 & TYC 6051-1123-1  \\
J100927.98$+$200519.4 & 1.395 & 12.3 &                  \\
J102122.83$-$284139.6 & 0.901 & 11.1 & TYC 6631-538-1   \\
J132357.01$+$433555.4 & 0.795 &  9.4 & EL~CVn, A1\,V    \\
J142951.60$-$244327.4 & 2.173 & 10.8 & HD 127051, F0    \\
J162545.15$-$043027.9 & 1.526 & 10.4 & HD 148070, A0    \\
J162842.31$+$101416.7 & 0.720 & 12.9 &                  \\
J181417.43$+$481117.0 & 1.798 & 10.7 & TYC~3529-1515-1\\
J204746.49$+$043602.0 & 1.563 & 11.8 & TYC 520-1173-1   \\
J210129.26$-$062214.9 & 1.290 & 11.5 & TYC 5204-1575-1  \\
J224905.95$-$693401.8 & 1.162 & 12.5 & TYC 9337-2511-1  \\
J232812.74$-$395523.3 & 0.768 & 13.3 &                  \\
\hline
\end{tabular}
\end{table}

\section{Target selection}
 
 The WASP survey is described in \citet{2006PASP..118.1407P} and
\citet{2008ApJ...675L.113W}. The survey obtains images of the night sky using
two arrays of 8 cameras each equipped with a charge-coupled device (CCD),
200-mm f/1.8 lenses and a broad-band filter (400\,--\,700\,nm).  Two
observations of each target field are obtained every 5\,--\,10 minutes using
two 30\,s exposures. The data from this survey are automatically processed and
analysed in order to identify stars with lightcurves that contain transit-like
features that may indicate the presence of a planetary companion. Lightcurves
are generated using synthetic aperture photometry with an aperture radius of
48\,arcsec. Data for this study were obtained between 2004 May 05 and 2011 Aug
02.

 We identified stars with lightcurves similar to WASP\,0257$-$25 in the WASP
photometric archive by inspection of the lightcurves folded on orbital periods
identified by the various transit detection algorithms used by this survey. We
looked for a characteristic lightcurve in which the primary eclipse has a
``boxy'' appearance, i.e., steep ingress and egress and a well defined flat
section,  and the secondary eclipse due to the transit of the pre-He-WD  has a
comparable depth but is shallower than the total eclipse caused by the
occultation. Stars were selected for inspection based on various combinations
of parameters that characterised the phase-folded lightcurve, e.g, the depth
and width of the eclipse. 

 For each star with this type of lightcurve we used a least-squares fit of a
lightcurve model and inspection of the available catalogue photometry similar
to the final analysis described below to verify that the primary eclipse is
due to the total eclipse by an A-type or F-type dwarf star of a smaller,
hotter star. We found 17 stars that satisfied these criteria, only one of
which (EL~CVn) has been previously catalogued as an eclipsing binary star.
Although EL~CVn was not previously known to contain a pre-He-WD, it is
currently the brightest known example of this type of binary star and is the
only one to-date with an entry in the General Catalogue of Variable Stars
\citep{2008IBVS.5863....1K}. For these reasons, and as a matter of
convenience, we  choose EL~CVn as the prototype of this class of variable
star. The new EL~CVn-type binary stars identified are listed in
Table~\ref{startable}.

 The new EL~CVn stars presented here are not a complete survey of the WASP
archive for this type of binary star.  The selection effects that affect this
incomplete survey of the WASP archive may be large and are difficult to
quantify. For example, WASP\,0845$+$53 was included in this study because
the lead author noticed a print-out of its lightcurve on a co-author's desk.

\section{Additional Observations}

\subsection{PIRATE photometry}
 Observations of WASP\,1628+10 were obtained with the 0.4-m PIRATE telescope
\citep{2011PASP..123.1177H} using the R filter and an exposure time of 30\,s
on various nights from 2012 June 18 to 2012 August 14. The facility, located
at the Observatorio Astron\`{o}mico de Mallorca (OAM) at 200m altitude, was
set up in its PIRATE~Mk~II configuration. This comprises a 17-inch PlaneWave
CDK telescope with focal length 2939-mm and an SBIG STX-16803 camera with
KAF-16803 CCD detector (4096$\times$4096, 9$\mu$m pixels), providing a
43$^{\prime}\times$×43$^{\prime}$ field of view. The camera is
thermo-electrically cooled and was operated at $-$20$^{\circ}$C. All frames
were taken with the Baader R filter; this has a performance similar to the
Astrodon Sloan r' filter used by the APASS survey \citep{2010SASS...29...45S}.
The data sets were calibrated in the standard way using flat field, dark and
bias frames. Synthetic aperture photometry of the target relative to 5
comparison stars was performed using the software package Maxim~DL  v.~5.18.
The aperture radius, ($\approx 4$ arc seconds), was set to twice the
full-width at half maximum of the stellar profiles in each image. 

\subsection{Spectroscopy}

 Spectroscopy of WASP\,0845$+$53 was obtained with the ISIS spectrograph on
the 4.2-m William Herschel Telescope (WHT) at the Observatorio del Roque de
los Muchachos on the nights 2010 April 23\,--\,26. We used the R600B grating
on the blue arm to obtain spectra covering the wavelength range
3650\,--\,5100\AA\ at a dispersion of 0.44\AA/pixel.   The exposure time was
600\,s and the signal-to-noise ratio in a typical spectrum as approximately
150 per pixel. Spectra were extracted from the images using the software
package {\sc pamela} and the spectra were analysed using the program {\sc
molly}.\footnote{\it deneb.astro.warwick.ac.uk/phsaap/software}
Observations of the star were bracketed by calibration arc observations and
the wavelength calibration interpolated to the time of mid-observation for
each spectrum. The RMS residual of a 4$^{\rmn{th}}$ order polynomial fit to
the 33 arc lines used for the wavelength calibration was typically 0.02\AA.
The resolution of the spectra is approximately 1.1\AA.

 Spectroscopy of WASP\,0843$-$11 and WASP\,1323+43 was obtained with the
Sandiford Cassegrain Echelle Spectrometer (SES) on the 2.1-m  Otto Struve
Telescope at McDonald Observatory on the nights 2012 January 04\,--\,16. The
exposure time used was 1800\,s. The mean dispersion is 0.037\AA/pixel and the
resolving power of the spectrograph is $R\approx 60\,000$. The typical
signal-to-noise per pixel is approximately 25 for WASP\,1323+43 and 5 for
WASP\,0843$-$11. These spectra cover the wavelength range 4011\,--\,4387\AA.
The spectra were reduced using {\sc iraf}.\footnote{{\sc iraf} is distributed
by the National Optical Astronomy Observatory, which is operated by the
Association of Universities for Research in Astronomy (AURA) under cooperative
agreement with the National Science Foundation.}

 Spectroscopy of WASP\,1323+43, WASP\,1625$-$04, WASP\,1628+10 and
WASP\,2101$-$06 was obtained using the Twin spectrograph on the 3.5-m
telescope at the Calar Alto observatory. We used the T12 600 line/mm grating
in the blue arm to obtain  spectra covering the approximate wavelength range
3290\,--\,5455\AA\ at a dispersion of 1.1\AA/pixel. Spectra were extracted
from the images using the software package {\sc pamela} and the spectra were
analysed using the program {\sc molly}. The resolution of the spectra is
approximately 2.3\AA. The RMS residual of a 3$^{\rmn{rd}}$ order polynomial
fit to the 15 arc lines used for the wavelength calibration was typically
0.65\AA.

\section{Analysis}

\begin{table*}
 \caption{Parameters for the lightcurve models fit by least-squares. $L_B/L_A$
is the luminosity ratio in the band noted, other  parameter definitions are
given in the text. $T_0$ is given as HJD(UTC)-2450000 and  $N$ is the number of
observations. See section \ref{syserrsec} for a discussion of possible
systematic errors in these parameters.}
\label{lcfitTable}
 \begin{tabular}{@{}lrrrrrrrrrrrr}
\hline
Star &
\multicolumn{1}{c}{${\rmn{T}}_0$} &
\multicolumn{1}{c}{P[d]} & 
\multicolumn{1}{c}{$J$}& 
\multicolumn{1}{c}{$i$}& 
\multicolumn{1}{c}{$s$}& 
\multicolumn{1}{c}{$k$}& 
\multicolumn{1}{c}{$x_A$}&
\multicolumn{1}{c}{$q$}& 
\multicolumn{1}{c}{$R_{\rmn{A}}/a$}& 
\multicolumn{1}{c}{$R_{\rmn{B}}/a$}& 
\multicolumn{1}{c}{$L_B/L_A$}&
\multicolumn{1}{c}{$N$}\\
Band &
\multicolumn{1}{c}{$\pm$} &
\multicolumn{1}{c}{$\pm$} &
\multicolumn{1}{c}{$\pm$} &
\multicolumn{1}{c}{$\pm$} &
\multicolumn{1}{c}{$\pm$} &
\multicolumn{1}{c}{$\pm$} &
\multicolumn{1}{c}{$\pm$} &
\multicolumn{1}{c}{$\pm$} &
\multicolumn{1}{c}{$\pm$} &
\multicolumn{1}{c}{$\pm$} &
\multicolumn{1}{c}{$\pm$} &
\multicolumn{1}{c}{RMS}\\
\hline
WASP\,0131$+$28 & 4053.3730 & 1.882752  & 1.239 & 86.5 & 0.2710 & 0.2904 &0.34& 0.075 & 0.2100 & 0.0610 & 0.1058 & 13590 \\
WASP       &         0.0003 & 0.000001  & 0.014 &  0.4 & 0.0025 & 0.0017 &0.04& 0.045 & 0.0019 & 0.0007 & 0.0011 & 0.010 \\
\noalign{\smallskip}                                                                                    
WASP\,0346$-$21 & 5178.3330 & 0.9285752 & 2.749 & 79.1 & 0.4278 & 0.1720 &0.40& 0.165 & 0.3650 & 0.0628 & 0.0800 & 22163 \\
WASP            &    0.0002 & 0.0000003 & 0.073 &  0.5 & 0.0048 & 0.0021 &0.09& 0.035 & 0.0042 & 0.0009 & 0.0007 & 0.008 \\
\noalign{\smallskip}                                                                                    
WASP\,0358$-$31 & 4148.9483 & 2.189309  & 3.35  & 84.5 & 0.2944 & 0.1294 &0.67& 0.119 & 0.2606 & 0.0337 & 0.0559 & 6770 \\
WASP            &    0.0009 & 0.000008  & 0.14  &  1.6 & 0.0094 & 0.0029 &0.14& 0.021 & 0.0082 & 0.0014 & 0.0007 & 0.008 \\
\noalign{\smallskip}                                                                                    
WASP\,0843$-$11 & 4846.8921 & 0.792833  & 2.468 & 75.6 & 0.5129 & 0.1591 &0.48& 0.176 & 0.4425 & 0.0704 & 0.0602 & 5070  \\
WASP         &       0.0003 & 0.000005  & 0.087 &  0.8 & 0.0074 & 0.0028 &0.10& 0.012 & 0.0064 & 0.0015 & 0.0010 & 0.007 \\
\noalign{\smallskip}                                                                                    
WASP\,0845$+$53 & 4808.7689 & 0.844143  & 2.45  & 90.0 & 0.3423 & 0.1311 &0.48& 0.219 & 0.303  & 0.0397 & 0.0415 & 11689 \\
WASP       &         0.0004 & 0.000003  & 0.22  &  3.3 & 0.0156 & 0.0058 &0.22& 0.034 & 0.013  & 0.0028 & 0.0019 & 0.026 \\
\noalign{\smallskip} 
WASP\,0939$-$19 & 5568.0975 & 1.0731807 & 2.70  & 75.4 & 0.5159 & 0.1515 &0.44& 0.144 & 0.4480 & 0.0679 & 0.0592 &28257 \\
WASP       &         0.0005 & 0.0000008 & 0.13  &  0.9 & 0.0084 & 0.0041 &0.13& 0.017 & 0.0075 & 0.0019 & 0.0013 &0.017 \\
\noalign{\smallskip}                                                                                    
WASP\,1009$+$20 & 4075.5706 & 1.39442   & 1.50  & 83.0 & 0.423  & 0.2188 &0.43& 0.135 & 0.347  & 0.0759 & 0.0707 & 3871 \\
WASP       &         0.0009 & 0.00001   & 0.11  &  2.9 & 0.018  & 0.0084 &0.21& 0.032 & 0.014  & 0.0046 & 0.0030 & 0.022 \\
\noalign{\smallskip}                                                                                    
WASP\,1021$-$28 & 4159.5203 & 0.9008980 & 2.447 & 82.1 & 0.4474 & 0.2070 &0.47& 0.143 & 0.3707 & 0.0767 & 0.1026 & 15912 \\
WASP       &         0.0003 & 0.0000002 & 0.050 &  0.6 & 0.0041 & 0.0021 &0.06& 0.027 & 0.0034 & 0.0010 & 0.0011 & 0.011 \\
\noalign{\smallskip}                                                                                    
WASP\,1323$+$43 & 4230.5558 & 0.795629  & 2.20  & 80.3 & 0.410  & 0.1856 &0.52& 0.192 & 0.346  & 0.0641 & 0.0749 &  5145 \\
WASP       &         0.0005 & 0.000008  & 0.12  &  1.8 & 0.014  & 0.0053 &0.20& 0.039 & 0.012  & 0.0030 & 0.0023 & 0.009 \\
\noalign{\smallskip}                                                                                    
WASP\,1429$-$24 & 4287.0364 & 2.173523  & 2.367 & 80.9 & 0.3672 & 0.2337 &0.49& 0.136 & 0.2977 & 0.0696 & 0.1284 & 14345 \\
WASP       &         0.0007 & 0.000002  & 0.066 &  0.5 & 0.0058 & 0.0039 &0.11& 0.041 & 0.0046 & 0.0016 & 0.0025 & 0.015 \\
\noalign{\smallskip}                                                                                    
WASP\,1625$-$04 & 4973.3928 & 1.5263234 & 2.081 & 82.5 & 0.2786 & 0.1702 &0.50& 0.126 & 0.2381 & 0.0405 & 0.0600 & 30610 \\
WASP       &         0.0003 & 0.0000009 & 0.037 &  0.3 & 0.0029 & 0.0016 &0.06& 0.026 & 0.0025 & 0.0005 & 0.0005 & 0.008 \\
\noalign{\smallskip}                                                                                    
WASP\,1628$+$10 & 4921.8532 & 0.7203633 & 1.794 & 73.7 & 0.4878 & 0.2270 &0.84& 0.097 & 0.3976 & 0.0902 & 0.0935 &30131  \\
WASP       &         0.0005 & 0.0000009 & 0.088 &  0.6 & 0.0083 & 0.0044 &0.16& 0.019 & 0.0066 & 0.0023 & 0.0026 &0.042  \\
\noalign{\smallskip}                                                                                    
WASP\,1628$+$10 & 6100.3682 & 0.7203633 & 1.90  & 72.9 & 0.491  & 0.209  &0.31& 0.092 & 0.4057 & 0.0849 & 0.0826 &  480  \\
PIRATE     &         0.0004 & (fixed)   & 0.17  &  0.8 & 0.010  & 0.009  &0.40& 0.022 & 0.0098 & 0.0029 & 0.0024 &  0.006\\
\noalign{\smallskip}                                                                                    
WASP\,1814$+$48 & 4717.0636 & 1.7994305 & 3.000 & 90.0 & 0.2813 & 0.0964 &0.59& 0.134 & 0.2565 & 0.0247 & 0.0277 & 48841\\
WASP            &    0.0001 & 0.0000005 & 0.064 &  1.3 & 0.0028 & 0.0010 &0.05& 0.015 & 0.0025 & 0.0004 & 0.0002 & 0.007 \\
\noalign{\smallskip}                                                                                    
WASP\,2047$+$04 & 4797.7220 & 1.563143  & 1.257 & 80.8 & 0.4994 & 0.2379 &0.50& 0.102 & 0.4034 & 0.0960 & 0.0709 & 30979 \\
WASP       &         0.0002 & 0.000001  & 0.033 &  1.2 & 0.0088 & 0.0033 &0.06& 0.023 & 0.0066 & 0.0024 & 0.0014 & 0.014 \\
\noalign{\smallskip}                                                                                    
WASP\,2101$-$06 & 4971.4605 & 1.2908592 & 2.205 & 89.4 & 0.2940 & 0.1470 &0.52& 0.130 & 0.2543 & 0.0373 & 0.0473 & 29238\\
WASP            &    0.0002 & 0.0000008 & 0.064 &  1.6 & 0.0090 & 0.0038 &0.07& 0.026 & 0.0067 & 0.0024 & 0.0017 & 0.013\\
\noalign{\smallskip}                                                                                    
WASP\,2249$-$69 & 5408.1882 & 1.162553  & 1.386 & 79.2 & 0.5609 & 0.2530 &0.51& 0.148 & 0.4477 & 0.1133 & 0.0862 & 18914 \\
WASP       &         0.0005 & 0.000003  & 0.040 &  1.2 & 0.0094 & 0.0033 &0.09& 0.026 & 0.0071 & 0.0026 & 0.0020 & 0.032 \\
\noalign{\smallskip}                                                                                    
WASP\,2328$-$39 & 4681.9316 & 0.7687015 & 1.940 & 81.6 & 0.4429 & 0.2984 &0.49& 0.252 & 0.3411 & 0.1018 & 0.1730 & 17086 \\
WASP       &         0.0004 & 0.0000003 & 0.079 &  0.9 & 0.0088 & 0.0057 &0.15& 0.074 & 0.0067 & 0.0027 & 0.0051 & 0.052 \\
\noalign{\smallskip}                                                                                    
\hline
\end{tabular}
\end{table*}
\begin{table*}
 \caption{Effective temperature estimates based on fitting the observed flux
distribution and the surface brightness ratio from the fit to the WASP
lightcurve. $N_{\rmn{f}}$ is the number of flux measurements included in our
least-squares fits.  We have assumed an error in E(B$-$V) of $\pm 0.034$ and
accounted for this additional uncertainty in the error estimates quoted for
T$_{\rmn{eff,A}}$ and T$_{\rmn{eff,B}}$.
\label{TeffTable}}
 \begin{tabular}{@{}lrrrrrrl}
\hline
Star &  E(B$-$V) &
\multicolumn{1}{l}{T$_{\rmn{eff,A}}$} & 
\multicolumn{1}{l}{T$_{\rmn{eff,B}}$} & 
\multicolumn{1}{l}{$\chi^2$} &
\multicolumn{1}{l}{$\chi_J^2$} &
\multicolumn{1}{l}{$N_{\rmn{f}}$} & Notes\\
\hline
WASP\,0131$+$28 & 0.069&$ 9500\pm700 $&$ 10500 \pm 2000$ &  5.5 & 0.00 &10 & a,b \\
WASP\,0346$-$21 & 0.057&$ 7400\pm200 $&$  9950 \pm ~400$ & 12.5 & 0.02 &12 & b,c \\
WASP\,0358$-$31 & 0.006&$ 7600\pm300 $&$ 12000 \pm 2000$ &  2.9 & 0.00 & 9 & c,d \\
WASP\,0843$-$11 & 0.043&$ 6900\pm150 $&$ ~9300 \pm ~200$ &  7.9 & 0.00 &12 &     \\
WASP\,0845$+$53 & 0.026&$ 8000\pm400 $&$ 15000 \pm 1800$ &  4.5 & 0.02 & 9 &     \\
WASP\,0939$-$19 & 0.063&$ 7150\pm250 $&$ 10150 \pm ~100$ &  3.2 & 0.13 & 8 &     \\
WASP\,1009$+$20 & 0.025&$ 8600\pm400 $&$ 10800 \pm ~700$ &  5.9 & 0.00 &10 &     \\
WASP\,1021$-$28 & 0.064&$ 7300\pm300 $&$ ~9800 \pm ~500$ &  0.2 & 0.00 & 8 & a   \\
WASP\,1323$+$43 & 0.021&$ 8250\pm350 $&$ 12000 \pm ~900$ &  6.3 & 0.00 & 9 & c   \\
WASP\,1429$-$24 & 0.088&$ 7150\pm200 $&$ ~9700 \pm ~300$ &  3.8 & 0.15 &14 & a   \\
WASP\,1625$-$04 & 0.226&$ 9500\pm500 $&$ 11500 \pm 1500$ &  4.6 & 0.00 &12 & b,c \\
WASP\,1628$+$10 & 0.056&$ 7200\pm300 $&$ ~9200 \pm ~250$ &  3.0 & 0.01 & 9 &     \\
WASP\,1814$+$48 & 0.039&$ 8000\pm300 $&$ 12500 \pm 1800$ &  8.0 & 0.00 &10 & a   \\
WASP\,2047$+$04 & 0.075&$ 8300\pm500 $&$ ~9000 \pm 1700$ &  3.5 & 0.00 & 8 &     \\
WASP\,2101$-$06 & 0.046&$ 9000\pm200 $&$ 13500 \pm 1000$ &  1.0 & 0.00 & 9 & a   \\
WASP\,2249$-$69 & 0.026&$ 7400\pm200 $&$ ~8800 \pm ~100$ &  4.2 & 0.65 &11 &     \\
WASP\,2328$-$39 & 0.017&$ 7500\pm450 $&$ ~9400 \pm ~450$ &  1.8 & 0.01 &11 &     \\
\hline
\noalign{\smallskip}
\multicolumn{8}{l}{a. GALEX NUV flux measurement(s) included in fit with low weight.} \\
\multicolumn{8}{l}{b. GALEX FUV flux measurement(s) included in fit with low weight.} \\
\multicolumn{8}{l}{c. GALEX NUV flux measurement(s) excluded from fit.} \\
\multicolumn{8}{l}{d. GALEX FUV flux measurement(s) excluded from fit.} 
\end{tabular}
\end{table*}

\begin{figure*}
\mbox{\includegraphics[width=0.99\textwidth]{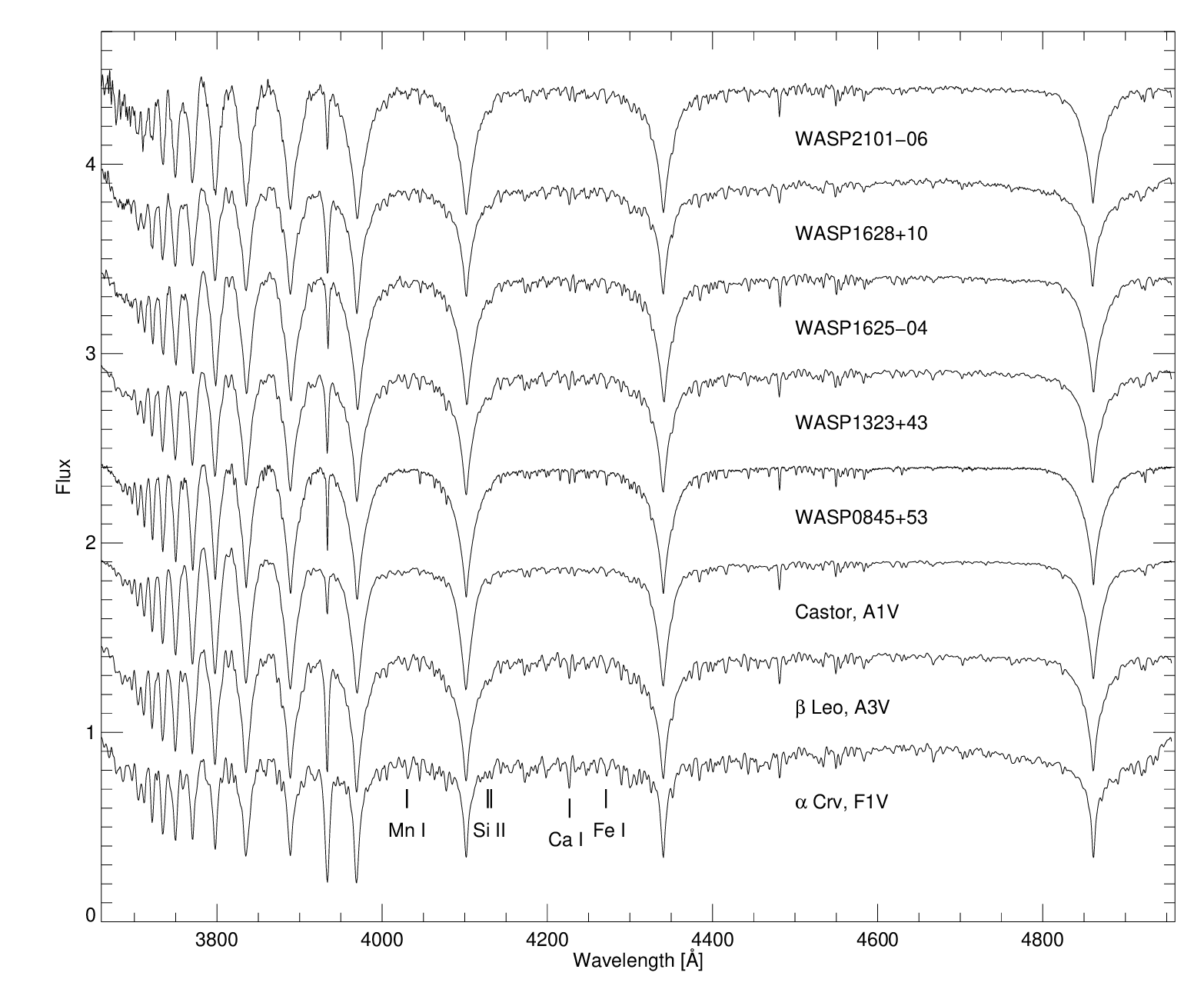}}
\caption{
Spectra of 5 EL~CVn-type binary stars compared to three stars of known
spectral type, as labelled \citep{2006MNRAS.371..703S}.  The spectra are
offset by multiples of 0.5 units for clarity. The Si\,II feature is enhanced
in Ap- and Am-type stars. The other spectral features indicated are useful for
assigning spectral type for A-type stars.
\label{spec}}
\end{figure*}

\subsection{Lightcurve analysis}
 We used {\sc jktebop}\footnote{\it
www.astro.keele.ac.uk/$\sim$\it jkt/codes/jktebop.html}
(\citealt{2010MNRAS.408.1689S} and references therein) to analyse the WASP and
PIRATE lightcurves using the {\sc ebop} lightcurve model
\citep{1981psbs.conf..111E,1981AJ.....86..102P}. The parameters of the model
are: the surface brightness ratio $J = S_B/S_A$, where $S_A$ is the surface
brightness of star A (``the primary star'') and similarly for star B (``the
secondary star'')\footnote{More precisely, {\sc jktebop} uses the surface
brightness ratio for the stars calculated at the centre of the stellar discs,
but for convenience we quote the mean surface brightness ratio here.}; the sum
of the radii relative to semi-major axis, $s=(R_{\rm A}+R_{\rm B})/a$; the ratio of the
radii, $k=R_{\rm B}/R_{\rm A}$; the orbital inclination, $i$;  the orbital period, $P$;
the UTC time (HJD) of the centre of the eclipse of star B, $T_0$; the mass
ratio,  $q=M_B/M_A$; the linear limb-darkening coefficient for star A, $x_A$.
We define star B  to be the smaller star in the binary in each case.  We
assumed that the orbit is circular and adopted a value $x_B=0.5$ for the
linear limb-darkening coefficient  of star B. Varying the value of $x_B$ has a
negligible effect on the lightcurve. \citet{2011A&A...529A..75C} have
calculated the gravity darkening coefficient, $y$, for stars in the Kepler pass-band,
which is similar to the WASP pass-band. The value of $y$ varies strongly and
non-monotonically over the effective temperature range of interest between the
values 0.1 to 0.8. The gravity darkening coefficient of the secondary star has
a negligible effect on the lightcurve so we set it to the value $y_{\rm
B}=0.5$. It is not possible to determine the gravity darkening coefficient of
the  primary star, $y_{\rm A}$, independently from the lightcurve because it
is strongly correlated with the value of $q$, so we include $y_{\rm A}$ as a
free parameter in the least-squares fit but limit this parameter to the range
$y_{\rm A} = 0.1$\,--\,0.8. We used the cyclic residual permutation method
(prayer-bead method) to estimate the standard errors on the lightcurve
parameters. The results are given in Table~\ref{lcfitTable}. The observed
lightcurves and the model fits are shown in Fig.~\ref{lcfitFig}.

\subsubsection{Systematic errors in lightcurve parameters\label{syserrsec}}

 The mass ratio estimated from the lightcurve depends on the amplitude of the
ellipsoidal effect, i.e., the variation between the eclipses caused by the
gravitational distortion of the stars, particularly the primary star. The
amplitude of the ellipsoidal effect also depends on the assumed rotation rate
for the primary star. We have assumed that the primary star rotates
synchronously, but this is known not to be the case in WASP\,0247$-$25.
Indeed, the mass ratio inferred from the lightcurve solution for
WASP\,0247$-$25 ($q=0.121 \pm 0.005$) is significantly different from the
correct value measured directly using spectroscopy ($q=0.137\pm 0.002$). A
similar discrepancy was observed by \citet{2012MNRAS.422.2600B} in the case of
KOI-74, an A-star with a low mass white dwarf companion in a 5.2-d orbit.

 In the case of WASP\,1628+10 we can compare the results derived from the
PIRATE photometry to the results derived from the WASP photometry in order to
estimate the likely level of systematic errors on the parameters.
WASP\,1628+10 is one of our fainter targets and there are several stars
1\,--\,2  magnitudes fainter within a few arc minutes of this star. It is
unlikely that these are bright enough to significantly affect the WASP
photometry.  The value of $s$ derived from the WASP lightcurve agrees well
with the value derived from the PIRATE lightcurve, as might be expected given
that this parameter is determined mainly by the duration of the eclipses.
However, there is disagreement just below the 2-$\sigma$ level in the value of
$k$ derived from the two lightcurves.  It is also possible that one or both of
the stars in WASP\,1628+10 is a pulsating star, as is the case for
WASP\,0247$-$25\,A (an SX~Phe-type star) and WASP\,0247$-$25\,B. The WASP
photometry is obtained over many pulsation cycles and so the effect of any
pulsations will be ``averaged-out'' in these lightcurves. The same does not
apply to the PIRATE lightcurve because some parts of the eclipses have only
been observed once or twice. So, if pulsations are present then there may be a
systematic error  in the value of $k$ derived from the PIRATE lightcurve.  In
principle, systematic errors in the estimation of the sky background level can
affect the measured amplitude of features in the lightcurve  such as the
eclipses and the ellipsoidal effect. In practice, the typical sky background
level in the WASP and PIRATE images is too low for this to be the main cause
of the discrepancy. The WASP photometry is processed using the {\sc sysrem}
algorithm \citep{2005MNRAS.356.1466T}. We inspected the power spectrum of the
corrections applied by this algorithm to the WASP lightcurve of WASP\,1628+10
and found that there is negligible power at the orbital frequency or its first
harmonic. Nevertheless, obtaining accurate photometry from wide-field images
such as those used in the WASP survey is not straightforward, so it is also
possible that the source of the discrepancy is some instrumental effect in the
WASP lightcurve. Whatever the source of this discrepancy may be, this level of
systematic error  in the value of $k$ is not large enough to affect our
conclusion that the secondary star in this binary system is a pre-He-WD. 

 In summary, there may be systematic errors in the values of $k$ and $q$
derived from the WASP lightcurves. In the case of WASP\,1628+10, the
systematic error in $R_{\rm A}/a$ is no more than a few per cent, but the
value of $R_{\rm B}/a$ may be in error by about 15\,per~cent.  This is not
sufficient to affect our conclusion that the secondary stars in these binary
systems are pre-He-WD.

\subsection{Effective temperature estimates.}
 We have estimated the effective temperatures of the  stars  by comparing the
observed flux distribution of each binary star to synthetic flux distributions
based on the BaSel 3.1 library of spectral energy distributions
\citep{2002A&A...381..524W}. Near-ultraviolet (NUV) and far-ultraviolet (FUV)
photometry was obtained from the GALEX GR6 catalogue\footnote{\it
galex.stsci.edu/GR6} \citep{2007ApJS..173..682M}. Optical photometry was
obtained from the NOMAD catalogue\footnote{\it www.nofs.navy.mil/data/fchpix}
\citep{2004AAS...205.4815Z}. Near-infrared photometry was obtained from the
2MASS\footnote{\it www.ipac.caltech.edu/2mass}  and DENIS\footnote{\it
cdsweb.u-strasbg.fr/denis.html} catalogues
\citep{2006AJ....131.1163S,2005yCat.2263....0T}.  The assumed surface gravity
has little effect on the model spectra and so we adopt nominal values of $\log
{\rmn{g}}_{\rmn{A}} = 4.0$ for the primary stars and $\log \rmn{g}_{\rmn{B}} =
5.0$ for the secondary stars. For the reddening we take the values for the
total line-of-sight reddening from the maps of
\citet{2011ApJ...737..103S}\footnote{\it
ned.ipac.caltech.edu/forms/calculator.html}. We compared the E(B$-$V) values
derived from these maps to the values derived using Str\"{o}mgren photometry
for 150 A-type stars by \citet{2004A&A...422..527S}. We find that standard
deviation of difference between these values is 0.034 magnitudes. This
additional source of error is accounted for in our analysis.

 We used linear interpolation to create a grid of spectral energy
distributions (SEDs) for the primary stars covering the effective temperature
range ${\rmn{T}}_{\rmn{eff,A}} = 6000$\,K -- 10\,000\,K in 25\,K steps,
10\,000\,K -- 13\,000\,K in 50\,K steps and 13\,000\,K -- 17\,000\,K in 100\,K
steps. We produced a similar grid for the secondary stars over the effective
temperature range  ${\rmn{T}}_{\rmn{eff,B}} = 7000$\,K\,--\,30\,000\,K.  Both
grids cover the metallicity range $[{\rmn{Fe/H}}] = -2.0 $ to $[{\rmn{Fe/H}}]
= 0.5$ in steps of 0.5. 

 For every pair of ${\rmn{T}}_{\rmn{eff,A}}$ and ${\rmn{T}}_{\rmn{eff,B}}$
values we integrated the synthetic SED of each star over a rectangular
band-pass that approximates the band-pass for each of the observed flux
measurements.  We then combined these synthetic flux distributions according
to the luminosity ratio of the stars in the WASP photometric band given in
Table~\ref{lcfitTable} and calculated the value of the scaling factor, $z$
that minimises the quantity \[ \chi_{f}^2 = \sum_{i=1}^{N_{\rmn{f}}}
\frac{(f_{{\rmn{obs}},i} - zf_{{\rmn{syn}},i})^2} {\sigma_i^2 +
(s_{\rmn{sys}}f_{{\rmn{obs}},i})^2},\] where $N_{\rmn{f}}$ is the number of
flux measurements included in the fit, $f_{{\rmn{obs}},i}$ are the observed
flux measurements, $f_{{\rmn{syn}},i}$ are the combined fluxes calculated from
the synthetic SEDs and $\sigma_i$ is the standard error on $f_{{\rmn{syn}},i}$
given in the source catalogue. The factor $s_{\rmn{sys}}$ is used to allow for
additional uncertainties in comparing model fluxes to observed fluxes, e.g.,
inconsistencies between the zero-point of the flux scale from different
catalogues. We included the surface brightness ratio measured from the WASP
lightcurve as an additional constraint in the least-squares fit by using the
figure of merit \[ \chi^2 = \chi^2_{f} + \chi^2_{J} = \chi^2_{f} +
\frac{(J_{\rmn{obs}}-J_{\rmn{syn}})^2}{\sigma_J^2 +
(s_{\rmn{sys}}J_{\rmn{obs}})^2}, \] where $J_{\rmn{obs}}$ and $\sigma_J$ are
taken from Table~\ref{lcfitTable} and $J_{\rmn{syn}}$ is the surface
brightness ratio in the WASP band calculated from the synthetic SEDs. In
combination with the fixed luminosity ratio this equivalent to adding the
ratio of the stellar radii as a constraint in the fit.

 There are five free parameters in this least-squares fit, $z$,
${\rmn{T}}_{\rmn{eff,A}}$, ${\rmn{T}}_{\rmn{eff,B}}$  and the value of [Fe/H]
for each star. We fit [Fe/H] independently for each star because the primary
stars may be chemically peculiar stars, e.g., Am-type stars,  and the
secondary stars may also have unusual surface chemical composition as a result
of their evolution or diffusion of elements in their atmospheres due to
gravitational settling and radiative levitation.  The values of [Fe/H] that
provide the best-fit to these broad-band flux measurements are unlikely to be
an accurate estimate of the stars' true surface composition and so we do not
quote them here. We have used the value $s_{\rmn{sys}}=0.05$ for all the fits
because this gives $\chi^2\approx N_{\rmn{f}}-N_{\rmn{par}}$, where
$N_{\rmn{par}}=5$ is the number of free parameters in the least-squares fit.
The saturation limits for the  GALEX instrument are not precisely defined  so
for measurements close to the nominal saturation limit we increased the
standard error by a factor of 10, rather than simply excluding measurements
near this limit. This was particularly useful in the case of WASP\,0131+28
where the GALEX fluxes were strongly under-predicted by the models if they were
completely excluded from the fit. The fits to the observed flux distributions
are shown in Fig.~\ref{fluxfitfig}. The values of ${\rmn{T}}_{\rmn{eff,A}}$,
${\rmn{T}}_{\rmn{eff,B}}$ derived are given in Table~\ref{TeffTable}. 

 We tested our method using the lightcurve solution for the WASP data of
XY~Cet by \citet{2011MNRAS.414.3740S}. Using the same method as described
above we derive effective temperatures ${\rmn{T}}_{\rmn{eff,A}}
=7865\pm610$\,K and  ${\rmn{T}}_{\rmn{eff,B}} = 7360\pm500$\,K. While these
values are not very precise, they are in good agreement with the values
${\rmn{T}}_{\rmn{eff,A}} = 7870\pm115$\,K and ${\rmn{T}}_{\rmn{eff,B}} =
7620\pm125$\,K derived from the analysis of the spectra of these stars. 

\subsection{Spectral types}
 We  have estimated the spectral types of our target stars by comparing the
hydrogen lines and other spectral
features\footnote{\it ned.ipac.caltech.edu/level5/Gray/frames.html} in  our low
resolution spectra to the spectra of bright stars with known spectral types
(Fig.~\ref{spec}). The spectral types derived are listed in
Table~\ref{sptytable}. These spectral types apply to the combined spectrum of
the binary. The optical spectrum is dominated by the light from star A and so
the spectral type of star A in each binary will be very similar to the value
given in Table~\ref{sptytable} unless star B has a very unusual spectrum. We
did not notice any obvious signs of chemical peculiarity in these spectra. The
Ca\,II K lines in some targets are slightly weak compared to those in the
standard stellar spectra, which may be an indication of chemical peculiarity,
but this line varies rapidly with spectral type so this is not a strong
indication in this case. In the two cases where spectral types are listed in
SIMBAD for our targets our spectral types are later than the published values
by two sub-types.

\begin{figure}
\mbox{\includegraphics[width=0.47\textwidth]{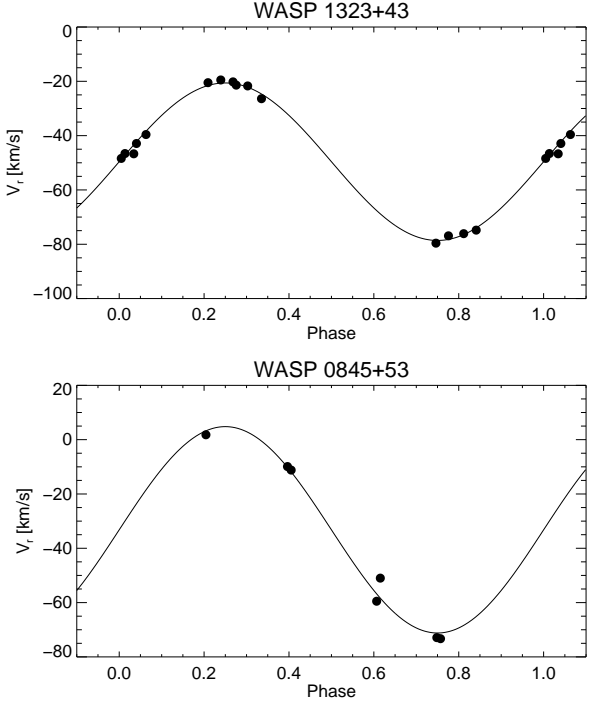}}
\caption{Measured radial velocities as a function of orbital phase for 
WASP\,1323+43 and WASP\,0845+53 (points) with  circular orbits fit by
least-squares (lines). \label{rvfitfig}} 
\end{figure}

\begin{table*}
\caption{Spectral types, radial velocities, proper motions
and distances for new EL CVn binaries. \label{sptytable} }
\begin{tabular}{@{}lllrrrrrrrrr}
\hline
Star & 
\multicolumn{2}{c}{Spectral type} &
\multicolumn{1}{l}{V$_r$} &
\multicolumn{1}{l}{$\mu_{\alpha}$} &
\multicolumn{1}{l}{$\mu_{\delta}$} &
\multicolumn{1}{l}{d} \\
& This paper & SIMBAD & 
\multicolumn{1}{l}{[\kms]} &
\multicolumn{1}{l}{[mas/y]} &
\multicolumn{1}{l}{[mas/y]} &
\multicolumn{1}{l}{[pc]} \\
\hline
WASP 0358$-$31&     &    A3&$   -2\pm 20$&$  1.4\pm1.0$&$ -3.5\pm1.0$&$ 535\pm 40$ \\
WASP 0843$-$11&     &      &$   24\pm 11$&$-11.1\pm1.0$&$  8.1\pm1.0$&$ 410\pm 30$ \\
WASP 0845$+$53&A2\,V&      &$  -42\pm  2$&$ -6.4\pm1.0$&$-46.1\pm1.6$&$1200\pm100$ \\
WASP 1323$+$43&A3\,V& A1\,V&$-58.6\pm0.3$&$ 19.5\pm0.5$&$-18.8\pm0.5$&$ 280\pm 20$ \\
WASP 1625$-$04&A2\,V&  A0  &$    9\pm  4$&$  6.3\pm1.0$&$  2.5\pm1.0$&$ 410\pm 30$ \\
WASP 1628$+$10&A2\,V&      &$  -59\pm 20$&$-14.1\pm1.1$&$  1.8\pm1.1$&$1040\pm 80$ \\
WASP 2101$-$06&A2\,V&      &$  -12\pm 16$&$ -4.1\pm1.0$&$-15.4\pm1.0$&$ 800\pm 60$ \\
\hline
\end{tabular}
\end{table*}

\subsection{Radial velocity measurements}
 We used a spectrum of the A4\,V star HD~145689 \citep{2003Msngr.114...10B}
as a template in a cross-correlation analysis of our spectra to measure the
radial velocities of the primary star. We analysed the spectral regions
4900\,--\,5300\,\AA, 4370\,--\,4830\,\AA\ and 4120\,--\,4320\,\AA\
independently for WASP\,0845+53 and took the mean radial velocity derived from
the peak of the cross correlation function. For WASP\,1323+43 we analysed the
entire spectrum excluding a 40\AA\ region around each Balmer line.  The radial
velocities derived and corrected for the radial velocity of the template taken
from the SIMBAD database ($-9$\,\kms) are listed in Table~\ref{rvtable}.  The
standard errors given in Table~\ref{rvtable} were found by requiring the
reduced chi-squared value of a circular orbit fit to be $\chi^2_r = 1$. The
parameters of the circular orbit fits are given in Table~\ref{rvfitTable} and
the fits are shown in Fig.~\ref{rvfitfig}.

 We attempted a similar analysis of the spectra obtained with the Twin
spectrograph but found that the radial velocities derived were only accurate
to about $\pm 20$\kms.  We obtained one or two spectra near each of the
quadrature phases, so these data  are sufficient to show that
WASP\,1625$-$04\,B, WASP\,1628+10\,B and WASP\,2101$-$06\,B are low mass
objects ($\la 0.3$\Msolar), but are not accurate enough to make a useful
estimate of these stars' masses. The radial velocities derived from the
McDonald spectra of WASP\,0843$-$11 are also affected by systematic errors due
to problems in combining data from different echelle orders in these low
signal-to-noise spectra. Again, we can confirm that WASP\,0843$-$11\,B is a
low mass star ($\la 0.3$\Msolar) but are not able to reliably estimate its
mass.

\begin{table}
\caption{Radial velocity measurements. 
\label{rvtable}
}
\begin{tabular}{@{}lrrl}
\hline
\multicolumn{1}{@{}l}{Star} 
& \multicolumn{1}{l}{HJD}
&\multicolumn{1}{l}{$V_r$}
& Source \\
\multicolumn{1}{l}{} 
& \multicolumn{1}{l}{$-2450000$}
&\multicolumn{1}{c}{[km\,s$^{-1}$]}
& \\
\hline
WASP\,0845$+$53 &  5310.3621 &$   -8.8 \pm  3.8$& WHT, ISIS \\
                &  5311.3687 &$  -18.9 \pm  3.8$& WHT, ISIS \\
                &  5311.3759 &$ -20.2 \pm  3.8$& WHT, ISIS \\
                &  5312.3903 &$ -68.5 \pm  3.8$& WHT, ISIS \\
                &  5312.3975 &$ -60.0 \pm  3.8$& WHT, ISIS \\
                &  5313.3539 &$ -81.9 \pm  3.8$& WHT, ISIS \\
                &  5313.3611 &$ -82.3 \pm  3.8$& WHT, ISIS \\
\noalign{\smallskip}
WASP\,1323$+$43 & 5937.9794  &$ -57.4 \pm  1.3$& McDonald \\
                & 5938.0028  &$ -55.7 \pm  1.3$& McDonald \\
                & 5938.0259  &$ -48.6 \pm  1.3$& McDonald \\
                & 5941.0084  &$ -85.1 \pm  1.3$& McDonald \\
                & 5941.0318  &$ -83.8 \pm  1.3$& McDonald \\
                & 5942.9687  &$ -30.4 \pm  1.3$& McDonald \\
                & 5942.9900  &$ -30.7 \pm  1.3$& McDonald \\
                & 5943.0161  &$ -35.4 \pm  1.3$& McDonald \\
                & 5930.9816  &$ -29.5 \pm  1.3$& McDonald \\
                & 5931.0052  &$ -28.5 \pm  1.3$& McDonald \\
                & 5931.0285  &$ -29.2 \pm  1.3$& McDonald \\
                & 5932.9999  &$ -88.6 \pm  1.3$& McDonald \\
                & 5933.0236  &$ -85.9 \pm  1.3$& McDonald \\
                & 5934.0083  &$ -55.6 \pm  1.3$& McDonald \\
                & 5934.0298  &$ -51.9 \pm  1.3$& McDonald \\
\hline
\end{tabular}
\end{table}

\begin{table}
 \caption{Parameters for least-squares fits of a circular orbit to our
measured radial velocities, $V_r = V_0 + K\sin((t-T_0)/P)$. The values of
$T_0$ and $P$ are taken from Table~\ref{lcfitTable} and $f_{\rm{m}}$  is the
mass function. 
\label{rvfitTable}
}
 \begin{tabular}{@{}lrrr}
\hline
Star &
\multicolumn{1}{l}{$V_0$} &
\multicolumn{1}{l}{K} &
\multicolumn{1}{l}{$f_{\rmn{m}}$}\\
&
\multicolumn{1}{l}{[\kms]} &
\multicolumn{1}{l}{[\kms]} &
\multicolumn{1}{l}{[\Msolar]}\\
\hline
WASP\,0845$+$53 &$ -42.2 \pm 1.5 $&$ 38.0 \pm 1.8 $ & 0.0048 $\pm $ 0.0007   \\
WASP\,1323$+$43 &$ -58.6 \pm 0.3 $&$ 29.0 \pm 0.4 $ & 0.0020 $\pm $ 0.0001   \\
\hline
\end{tabular}
\end{table}

\subsection{Kinematics}
 We have calculated the Galactic U, V, W velocity components for the stars in
our sample with measured radial velocities. For WASP\,0845+53 and
WASP\,1323+43 we use the centre-of-mass velocity and its standard error 
from Table~\ref{rvfitTable}. The radial velocity of WASP\,0358$-$31 is taken
from \citet{2011AJ....141..187S}, but rather than the quoted standard error we
use a nominal standard error of 20\kms\ to account for the unknown
contribution of the orbital velocity to this measurement. For the
other stars we use the mean radial velocity and its standard error measured
from the Twin spectra. For the proper motions of the stars we used the average
of the results from the  PPMXL, UCAC4 and NOMAD catalogues
\citep{2010AJ....139.2440R,2013AJ....145...44Z, 2004AAS...205.4815Z}. The
proper motion values in these catalogues are not independent so we assigned an
error of 1 mas/yr to these values or used the standard error of the mean if
this is larger. To estimate the distance we assumed that the primary star has
a mass in the range 1\,--\,2\Msolar\ and then used the values of $P$, $q$ and
$R_{\rmn{A}}/a$ from Table~\ref{lcfitTable} to estimate the radius of this
star. We then used the 2MASS apparent $K$-band magnitude corrected for the
contribution from the secondary star and the calibration
of $K$-band surface brightness as a function of effective temperature from
\citet{2004A&A...426..297K} to estimate the distance to the binary. The values
of U, V and W were calculated using the methods described in
\citet{2006A&A...447..173P} and are given in Table~\ref{uvwtable}, together
with  the eccentricity ($e$) and z-component of the angular momentum
(${\rmn{J}}_{\rmn{Z}}$)
of the systems' Galactic orbits.

\begin{figure}
\mbox{\includegraphics[width=0.47\textwidth]{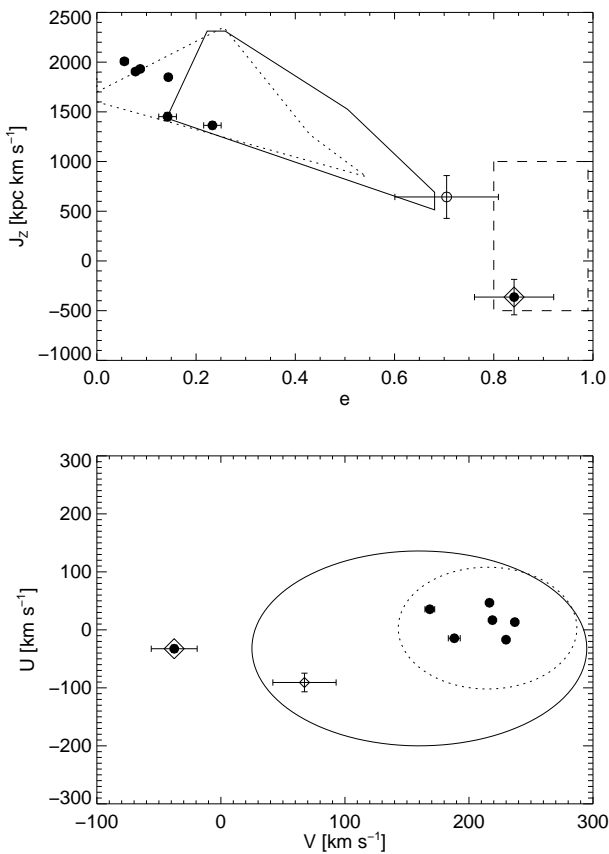}}
\caption{Upper panel: Eccentricity ($e$) and z-component of the angular
momentum (${\rmn{J}}_{\rmn{Z}}$) of the  Galactic orbits for our targets. The regions in which
thin-disk (dotted lines), thick-disk (solid lines) and halo stars
(dashed-lines) are found based on the results of \citet{2006A&A...447..173P}
are indicated. Lower panel: Galactic U and V velocities of our targets.
Contours are 3-$\sigma$ limits for the U-V distribution of main-sequence
thin-disk and thick-disk stars. In both panels the open symbol shows the
location of WASP\,0247$-$25  and WASP0845$+$53 is highlighted with an open
diamond symbol.
\label{uvplot}} 
\end{figure}

 Fig.~\ref{uvplot} shows our targets in the U-V and e-${\rmn{J}}_{\rmn{Z}}$
diagrams. WASP\,0845+53 is clearly a halo star. The e-${\rmn{J}}_{\rmn{Z}}$
diagram enables us to identify 5 thin-disk stars -- these are noted in
Table~\ref{uvwtable}. The remaining two targets are either thin-disk or
thick-disk  stars. We note in passing that the position of WASP\,0247$-$25 in
the e-${\rmn{J}}_{\rmn{Z}}$ diagram is consistent with the assumption
used in \citet{2013Natur.498..463M}  that this star belongs to the thick-disk
population.

\begin{table*}
\caption{Galactic U, V, W velocity components of the stars in our sample and
WASP\,0247$-$25. The eccentricity ($e$) and the z-component of the angular
momentum (${\rmn{J}}_{\rmn{Z}}$) of the stars'
Galactic orbits are also given. The population to which each star belongs has
been estimated based 
on the stars' positions in the U\,--\,V and $e$\,--\,${\rmn{J}}_{\rmn{Z}}$ diagrams.  
\label{uvwtable}
}
\begin{tabular}{@{}lrrrrrll}
\hline
Star &
\multicolumn{1}{c}{U} &
\multicolumn{1}{c}{V} &
\multicolumn{1}{c}{W} &
\multicolumn{1}{c}{e} &
\multicolumn{1}{c}{${\rmn{J}}_{\rmn{Z}}$}&
\multicolumn{2}{c}{Population} \\
&
\multicolumn{1}{c}{[\kms]} &
\multicolumn{1}{c}{[\kms]} &
\multicolumn{1}{c}{[\kms]} &
&
\multicolumn{1}{c}{[kpc \kms]} &
U\,--\,V & $e$\,--\,${\rmn{J}}_{\rmn{Z}}$ \\
\hline
WASP 0247$-$25&$ -91\pm 16$&$ 67\pm 26$&$ 34\pm 13$&$0.705 \pm 0.104$&$ 643\pm 215 $&thick disc & thick disc/halo  \\
WASP 0358$-$31&$  17\pm  2$&$219\pm  2$&$ 10\pm  2$&$0.077 \pm 0.007$&$ 1905\pm  17 $&disc & thin disc  \\
WASP 0843$-$11&$ -17\pm  2$&$230\pm  2$&$  1\pm  2$&$0.055 \pm 0.007$&$ 2008\pm  14 $&disc & thin disc  \\
WASP 0845$+$53&$ -33\pm  6$&$-38\pm 19$&$-31\pm  4$&$0.841 \pm 0.079$&$ -363\pm 178 $&halo & halo       \\
WASP 1323$+$43&$  47\pm  2$&$217\pm  1$&$ -7\pm  1$&$0.144 \pm 0.006$&$ 1848\pm   4 $&disc & thin disc  \\
WASP 1625$-$04&$  14\pm  1$&$251\pm  2$&$  2\pm  2$&$0.087 \pm 0.007$&$ 1931\pm  12 $&disc & thin disc  \\
WASP 1628$+$10&$ -15\pm  4$&$188\pm  5$&$ 62\pm  5$&$0.142 \pm 0.017$&$ 1451\pm  42 $&disc & thin disc\\
WASP 2101$-$06&$  35\pm  3$&$170\pm  4$&$  3\pm  3$&$0.232 \pm 0.017$&$ 1363\pm  33 $&disc & disc\\
\hline               
\end{tabular}
\end{table*}

\subsection{Masses and radii}
\begin{figure}
\mbox{\includegraphics[width=0.47\textwidth]{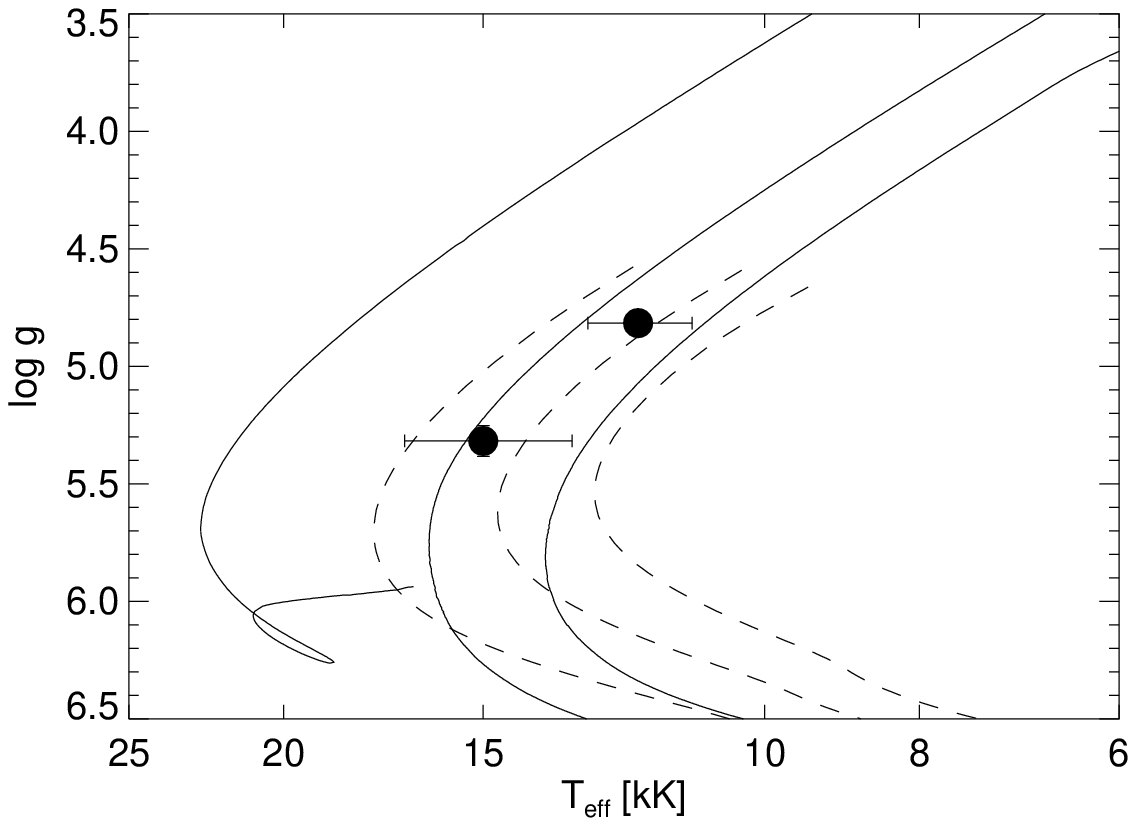}}
\caption{Effective temperature (\Teff) and surface gravity (\logg) of
WASP\,0845+53\,B ($\Teff=15\,000$\,K, $\logg=  5.3$) and WASP\,1323+43\,B
($\Teff=12\,000$\,K, $\logg = 4.8$) compared to models for the formation of
low-mass white dwarfs. Solid lines show (from left to right) the models of
\citet{1998A&A...339..123D} for masses 0.234\Msolar, 0.195\Msolar\ and
0.179\Msolar. Dashed lines show (from left to right) the models of
\citet{2002MNRAS.337.1091S} for Z=0.001 and masses 0.197\Msolar, 0.183\Msolar\
and 0.172\Msolar. \label{Tefflogg}} 
\end{figure}

 The surface gravity of the secondary star 
can be derived from the analysis of an SB1 binary with total eclipses without
any assumptions about the masses or radii of the stars
\citep{ 2004MNRAS.355..986S}. The effective
temperatures and surface gravities  of WASP\,0845+53\,B ($\logg=
5.32\pm0.07$) and WASP\,1323+43\,B ($\logg = 4.82\pm0.04$) are compared to
models for the formation of low-mass white dwarfs in Fig.~\ref{Tefflogg}.
These models predict that the masses of WASP\,0845+53\,B and WASP\,1323+43\,B
are both approximately 0.19\Msolar. 

\begin{figure}
\mbox{\includegraphics[width=0.47\textwidth]{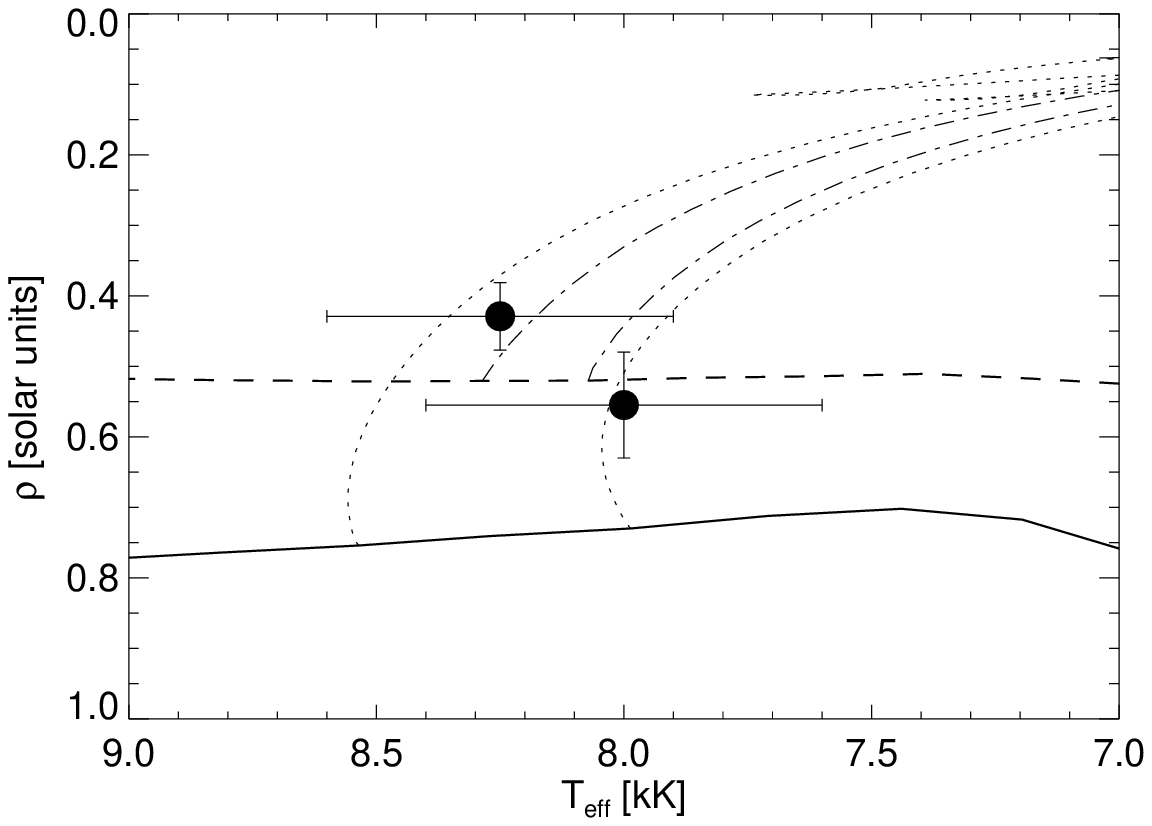}}
\caption{Mean density and effective temperature of WASP\,0845+53\,A
($\Teff=8\,000$\,K, $\rho=  0.56\rho_{\odot}$) and WASP\,1323+43\,A
($\Teff=8\,250$\,K, $\rho = 0.43\rho_{\odot}$) compared to the stellar models
of \citet{2008ApJS..178...89D}. Zero-age main sequence isochrones are shown
for models with $({\rmn{[Fe/H]}},[\alpha{\rmn{/Fe]}}) = (0.0, 0.0)$
(short-dashed line); $(-0.5, 0.2)$, (solid line). Evolutionary tracks are
plotted as follows: $(0.0,0.0)$, 1.70\Msolar\ and 1.75\Msolar\ (dash-dot
line); $(-0.5, 0.2)$, 1.45\Msolar\ and 1.5\Msolar\ (dotted lines).
 \label{TrhoFig}} 
\end{figure}

  The mean stellar density can be estimated from the parameters given in
Table~\ref{lcfitTable} via Kepler's Law using the equation $\rho/\rho_{\odot}
= 0.0134\left[(R_A/a)^3(P/d)^2(1+q)\right]^{-1}$. We compared the effective
temperature (\Teff) and mean stellar density ($\rho$) of WASP\,0845+53\,A and
WASP\,1323+43\,A to models from the Dartmouth Stellar Evolution Database
\citep{2008ApJS..178...89D}. A zero-age main sequence star (ZAMS) model with
solar composition and a mass of 1.75\Msolar\ provides a good match to
WASP\,1323+43\,A in the $\rho$\,--\,\Teff\ plane (Fig.~\ref{TrhoFig}). A ZAMS
model with the same composition but a slightly lower mass provides a good fit
for WASP\,0845+53\,A. Both stars will be less massive if they have a more
metal-poor composition, which may be more appropriate for the halo star
WASP\,0845+53. For example, WASP\,0845+53\,A is well matched by a model with a
mass of 1.45\Msolar\ and a composition $({\rmn{[Fe/H]}},[\alpha{\rmn{/Fe]}})
= (-0.5, 0.2)$ for an apparent age of about 1 Gyr. Note that this is the age
of a single star model, the binary system may be much older because the A-type
star will have gained mass from its companion during the formation of the
pre-He-WD.

\subsection{Notes on individual objects}
\begin{description}
\item[{\bf WASP\,0358$-$31}]{This star is listed in the 3$^{\rmn{rd}}$ data
release of the RAdial Velocity Experiment (RAVE) catalogue
\citep{2011AJ....141..187S}. The stellar parameters from this catalogue are
\Teff=7647\,K, \logg=4.17, [M/H]=0.07 and [$\alpha$/Fe]=0.00. This effective
temperature estimate is in  good agreement with our own estimate of the
primary star effective temperature based on the flux distribution of the star
and lightcurve solution (Table~\ref{TeffTable}). It should be noted that the
stellar parameters from the RAVE catalogue do not account for the contribution
of the secondary star to the observed spectrum. The distance to this star
estimated by \citet{2010A&A...522A..54Z} based on an analysis of the RAVE
spectrum ($504\pm182$) is in good agreement with the value in
Table~\ref{uvwtable}, though less precise.} 
\item[{\bf WASP\,1323+43}]{The estimated distance to this star given in
Table~\ref{uvwtable} is in good agreement with the parallax measured using the
Hipparcos satellite \citep{2007A&A...474..653V}. This star was detected as a
periodic variable star using the Hipparcos epoch photometry by
\citet{2002MNRAS.331...45K}, although the period measured from those data was
found to be incorrect by a factor of 2 by \citet{2004IBVS.5557....1O}. This
star appears in two libraries of stellar spectra as a standard A1V-type star
\citep{1984ApJS...56..257J,1992ApJS...81..865S}} 
\end{description}


\section{Discussion and conclusions}
 The positions of  WASP\,0845+53\,B and WASP\,1323+43\,B in
Fig.~\ref{Tefflogg} show that these are certainly pre-He-WD with masses
$\approx 0.19$\Msolar. The available spectroscopy for WASP\,1625$-$04,
WASP\,1628+10, WASP\,2101$-$06 and WASP\,0843$-$11 show that the secondary
stars in these binary systems have masses $\la 0.3$\Msolar\ and effective
temperature $\sim 10\,000$\,K, which is exactly as expected for pre-He-WDs.
 The minimum mass for a core helium burning star is about 0.33\Msolar, but
these anomolously low masses only occur for a narrow range of initial stellar
mass around 2.3\Msolar\ \citep{2009A&A...507.1575P, 2002MNRAS.336..449H}.
Although our upper limit on the mass of these stars does not completely
exclude the possibility that they are core helium burning stars, the following
argument does make this extremely unlikely, both for these stars and for the
stars for which we do not yet have any spectroscopy.

 The models of \citet{2009A&A...507.1575P} show that low-mass core helium
burning stars only have effective temperatures $\sim 10\,000$\,K
during phases when they are evolving very rapidly or when their luminosity is
$\log(L/\Lsolar)\ga 1$. If we assume that smaller star in the binary has a
luminosity $\log(L_{\rm B}/\Lsolar) = 1$ then the luminosity ratios in
Table~\ref{lcfitTable} and effective temperatures in Table~\ref{TeffTable}
require that the larger star in these binaries have radii
$R_A\approx5$\,--\,10\Rsolar. These radii can be combined with the value of
$R_A/a$ from Table~\ref{lcfitTable} and Kepler's Law to make an estimate of
the mass of star A in each binary.  We used bolometric corrections for the
V-band from \citet{1994MNRAS.268..119B} to convert the luminosity ratio in the
WASP band from Table~\ref{lcfitTable} to a bolometric luminosity ratio. We
find that the assumption $\log(L_{\rm B}/\Lsolar) = 1$ leads to masses from
$12\pm2$\Msolar\ to $115\pm16$\Msolar, with a typical value of
$40\pm10$\Msolar. This is clearly much too high for the mass of an A-type star
that can fit into an orbit with a period of less than 2.2\,days. 
Nevertheless, given the current observational uncertainties (particularly for
those stars without spectroscopic follow-up) it may be possible that a few of
these binary systems contain stars with a small carbon-oxygen core produced
during an evolutionary phase not explored by the models of
\citeauthor{2009A&A...507.1575P}. 

 WASP\,0845+53\,A is a blue-straggler, i.e., an apparently young star in an
old stellar population (the Galactic halo).  If we  assume that the red giant
progenitor of WASP\,0845+53\,B had a mass close to the halo turn-off mass
($\approx 0.8\Msolar$) and that this was initially the more massive star in
the binary system, then we see that WASP\,0845+53\,A must have gained   about
0.6\Msolar\ to get to a current mass of  $\approx 1.45\Msolar$. 

 With one exception (WASP\,1429$-$24, F0) the spectral types of the stars are
all in the range A0\,--\,A4. Dwarf stars with normal compositions in this
range of spectral type have $\Teff\approx$8000\,--\,9400\,K
\citep{2012ApJ...746..101B}. The effective temperatures we have estimated by
fitting the stars' flux distributions are generally within this range, but for
WASP\,1628+10 $\Teff$ is about 1000\,K cooler than this.  The spectral type
expected for this star based on our estimate of $\Teff$  is approximately A8.
Similarly, WASP0346$-$21, is expected to have a spectral type close to F0
based on our estimate of $\Teff$ but the published spectral type is A4\,IV. If
we assume that these stars have reddening E(B$-$V) approximately 0.15
magnitides larger than predicted by the reddening maps, as has observed for
some other A-type stars \citep{2004A&A...422..527S}, then we find that the
effective temperatures derived are in good agreement with the spectral types.
If there is a discrepancy between the effective temperatures derived from our
flux-fitting method and the spectral types then the atmospheric composition of
these stars may be very different to any of the compositions assumed for the
BaSel 3.1 library. The resolution and signal-to-noise of the spectra presented
here are not sufficient to explore this issue further. A detailed analysis of
high resolution spectra with good signal-to-noise would help us to better
understand this problem, particularly if the spectra can be obtained during
primary eclipse when there is no contribution from the companion star. In
addition, high resolution spectra covering the interstellar Na\,I D-lines can
be used to make independent estimates of the reddening to these stars
\citep{1997A+A...318..269M}.

 Until the discovery of WASP\,0247-25, very few pre-He-WD were known and none
of these were easy to study, being either faint, or with unseen companion
stars, or both \citep{2011MNRAS.418.1156M}. Our discovery of 17 new, bright
eclipsing binary systems containing these rarely observed stars opens up the
possibility of studying the formation of very low-mass white dwarfs in great
detail. These discoveries also show the great value of the WASP photometric
archive for the discovery and study of rare and interesting types of variable
star. 

 The large number of photometric observations and high cadence of the WASP
photometry makes it possible to identify the characteristic ``boxy'' primary
eclipse in the lightcurve. This feature combined with a shallower secondary
eclipse due to a transit is an unambiguous signal that a short period binary
star must contain a pre-He-WD or similar highly evolved star. It would not be
so straightforward to identify a pre-He-WD at an earlier phase of its
evolution when it is cooler than the dwarf star. Several binary systems have
been identified using Kepler photometry in which an A-type or B-type star has
a young, low-mass white dwarf companion, i.e., stars at a more advanced
evolutionary phase than the pre-He-WDs in EL~CVn-type binaries. The Kepler
binary systems have orbital periods in the range 2.6\,--\,23.9\,days
\citep{2012ApJ...748..115B, 2010ApJ...713L.150R,2011ApJ...728..139C}. This
suggests that there are likely to be more pre-He-WD awaiting discovery in the
WASP data, particularly at longer orbital periods. With a more systematic
approach to discovering these binaries it may be possible to learn more about
their formation by comparing the distributions of observed properties for a
more complete sample to the predictions of binary population synthesis models.

\section*{Acknowledgements}
 This research has made use of the SIMBAD database, operated at CDS,
Strasbourg, France. The research leading to these results has received funding
from the European Research Council under the European Community's Seventh
Framework Programme (FP7/2007--2013)/ERC grant agreement n$^\circ$ 227224
(PROSPERITY), as well as from the Research Council of the University of Leuven
under grant agreement GOA/2013/012. This work was supported by the Science and
Technology Facilities Council [grant numbers ST/I001719/1, ST/J001384/1]. SG
and DS were supported by the Deutsche Forschungsgemeinschaft (DFG) through
grants HE1356/49-1 and HE1356/62-1.

\bibliographystyle{mn2e}  
\bibliography{wasp}

\appendix
\section{Additional figures}

\begin{figure*}
\includegraphics[width=1.00\textwidth]{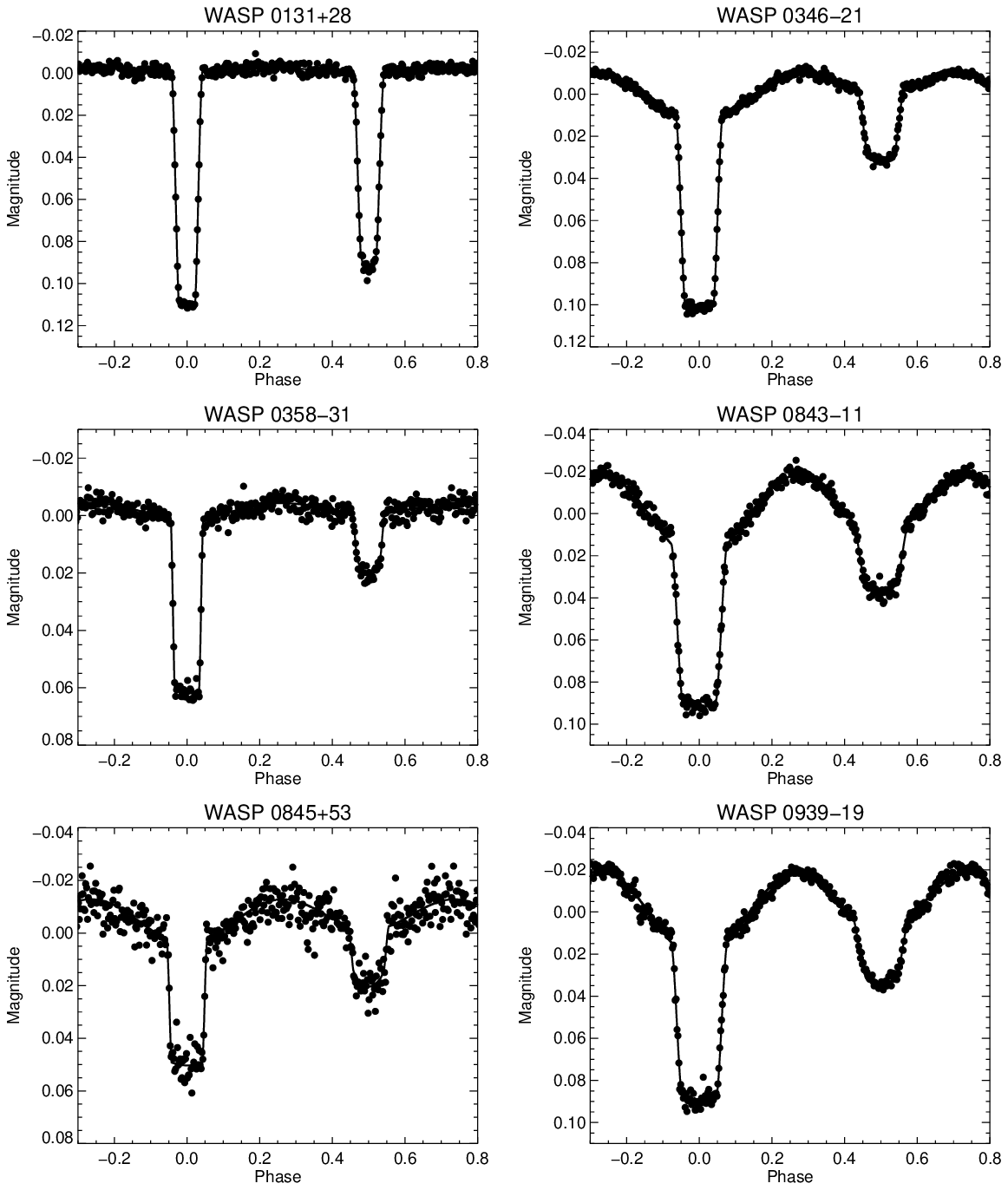}
\caption{ Observed lightcurves for our sample (points) with model lightcurves
fit by least-squares (lines).  The observed data have been binned in phase
(bin width 0.0025) for display purposes with the exception of the PIRATE data
for WASP\,1628+10.\label{lcfitFig}}
\end{figure*}

\begin{figure*}
\includegraphics[width=1.00\textwidth]{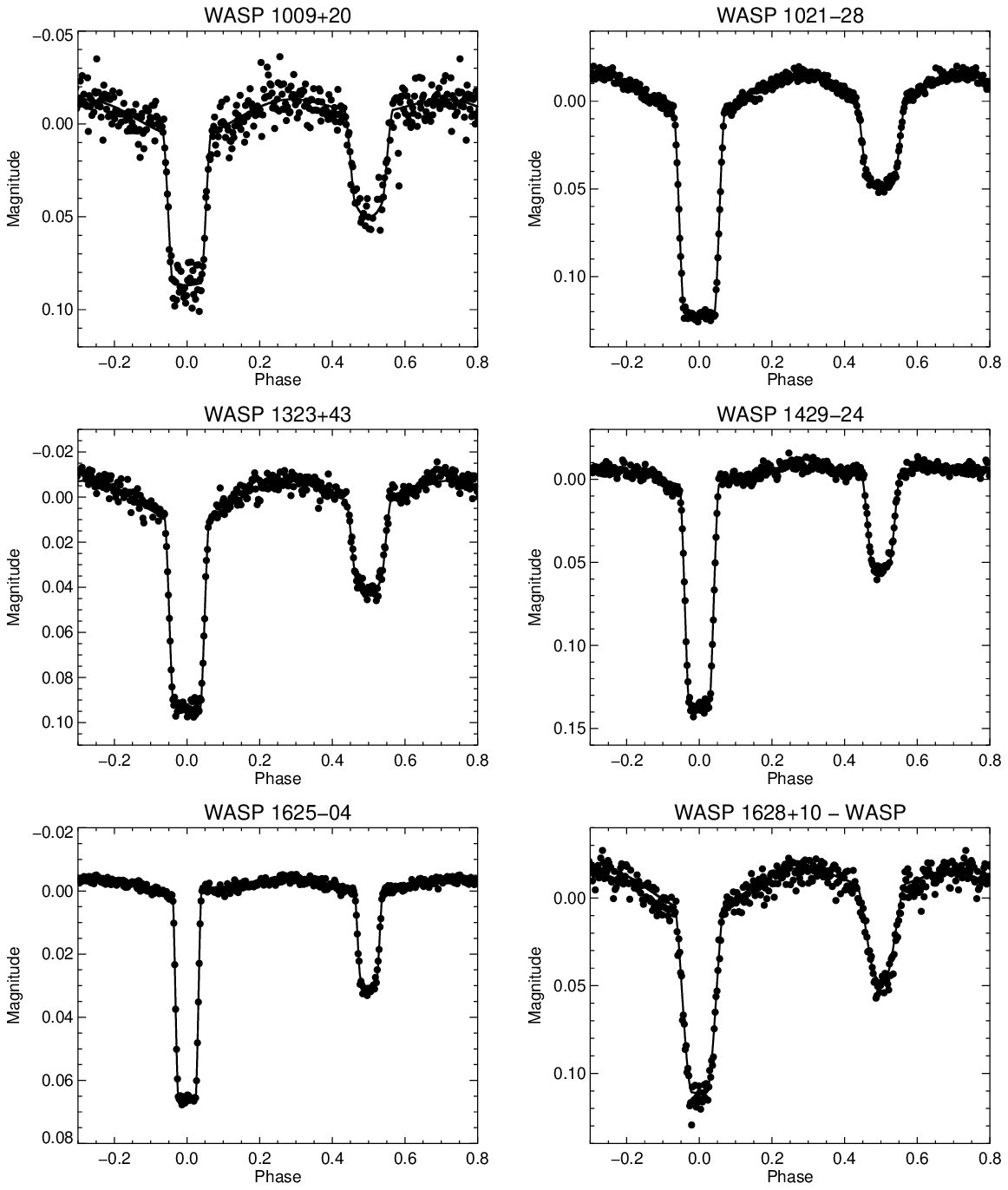}
\contcaption{}
\end{figure*}

\begin{figure*}
\includegraphics[width=1.00\textwidth]{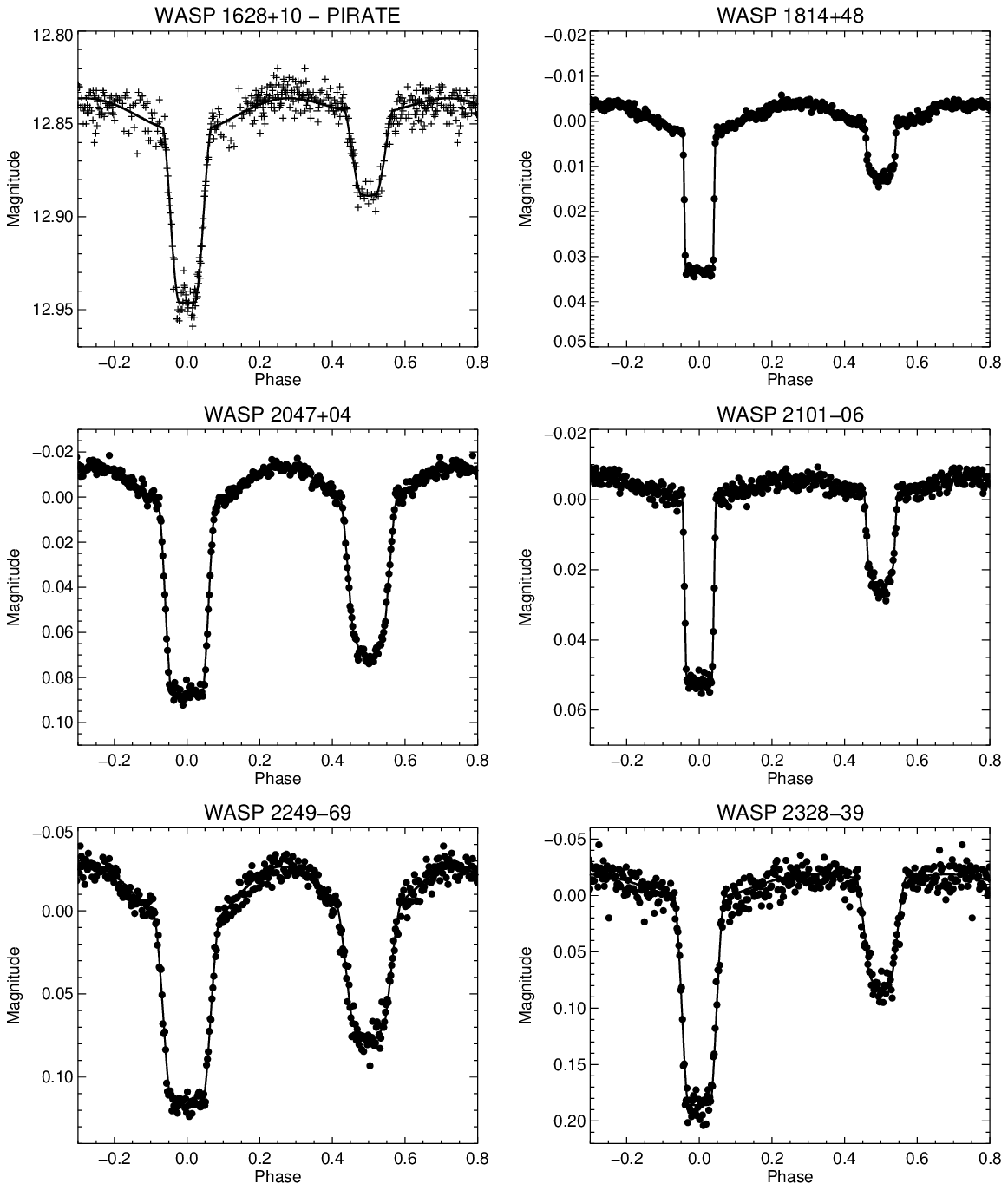}
\contcaption{}
\end{figure*}


\begin{figure*}
\mbox{\includegraphics[width=0.49\textwidth]{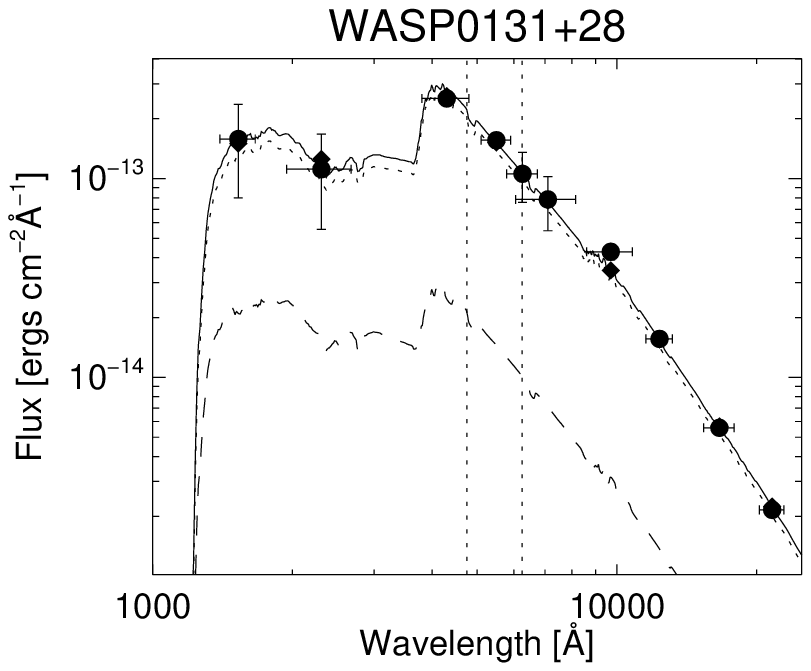}}
\mbox{\includegraphics[width=0.49\textwidth]{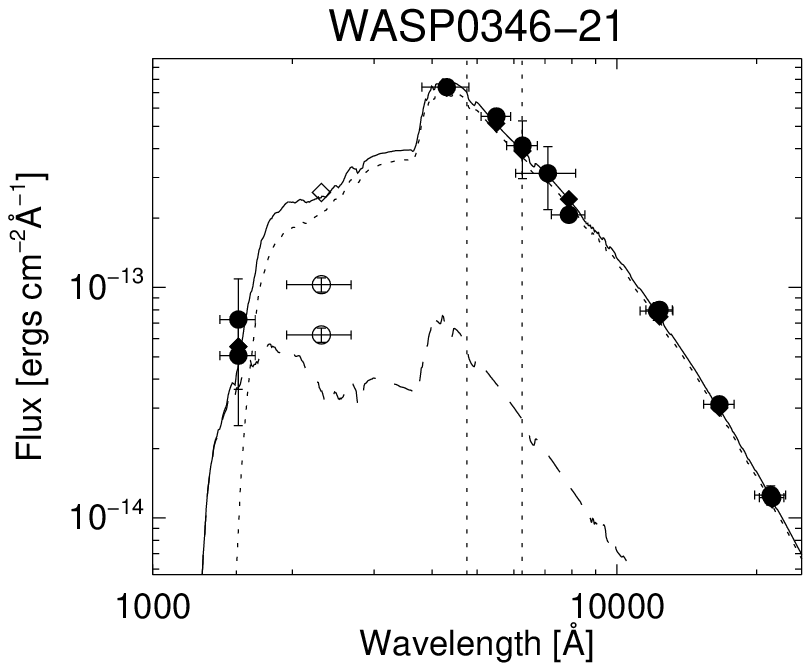}}
\mbox{\includegraphics[width=0.49\textwidth]{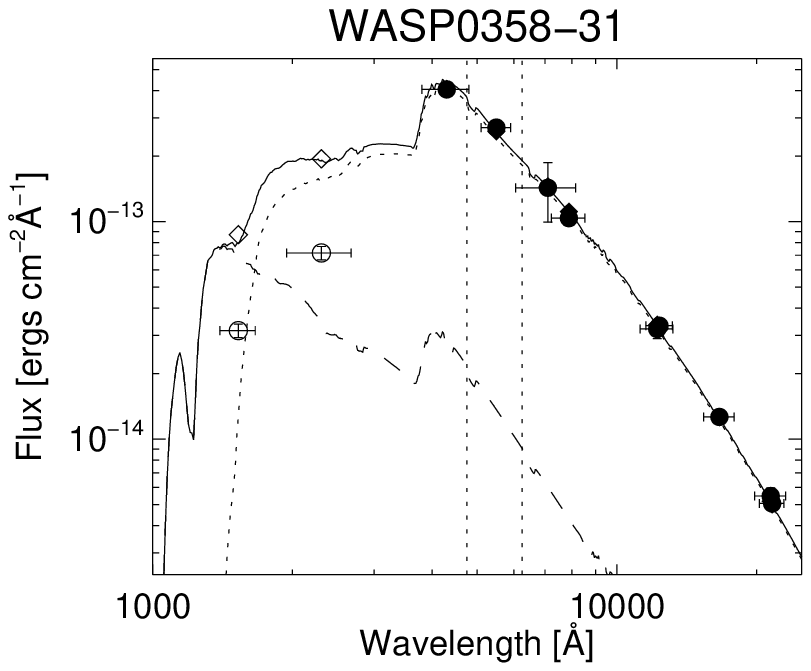}}
\mbox{\includegraphics[width=0.49\textwidth]{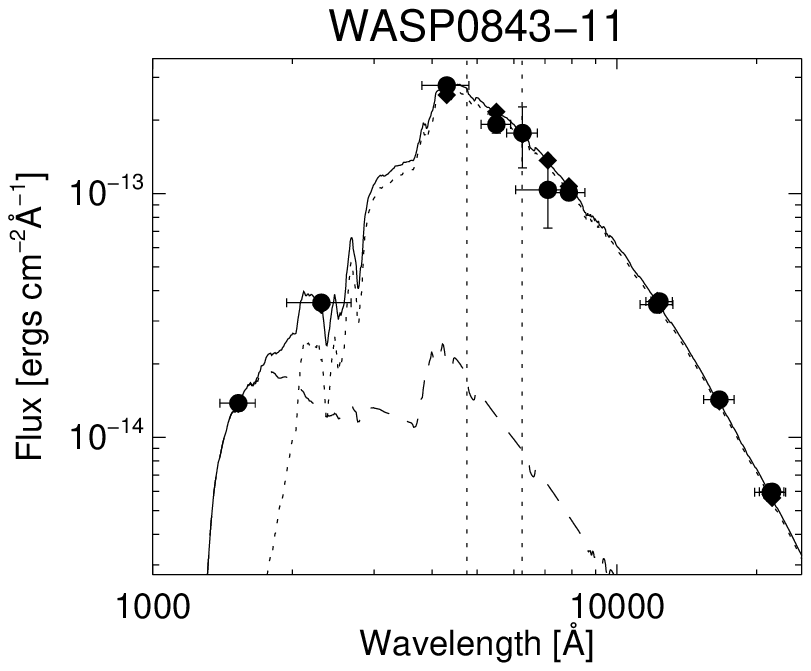}}
\mbox{\includegraphics[width=0.49\textwidth]{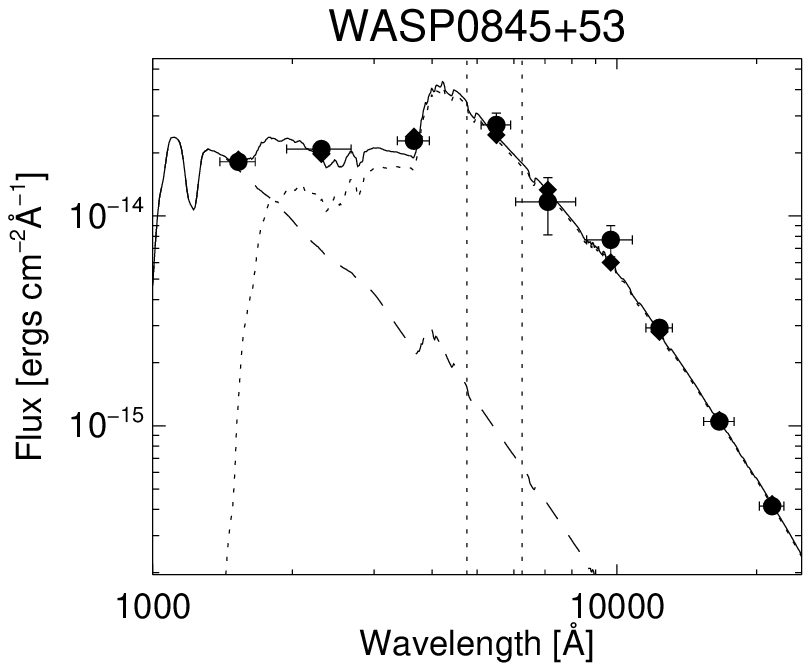}}
\mbox{\includegraphics[width=0.49\textwidth]{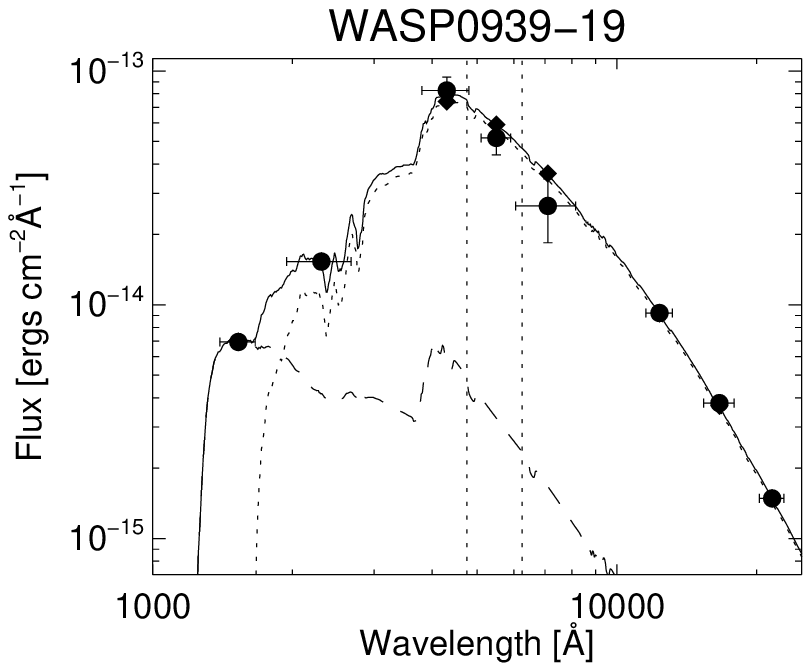}}
\caption{Model fits to the observed flux distributions used to estimate the
effective temperatures of the stars in our sample. The observed fluxes are
shown as circles with error bars. The predicted contributions of each star to
the observed fluxes for the effective temperatures given in
Table~\ref{TeffTable} are shown as dotted or dashed lines and their sum is
shown as a solid line. The models have been smoothed slightly for clarity in
these plots. Diamonds show the result of integrating the total model flux over
the band-width indicated by horizontal error bars on the observed fluxes. Only
data plotted with filled symbols were used in the least-squares fits. The
assumed band-width of WASP photometry is indicated with vertical dotted lines.
\label{fluxfitfig}} 
\end{figure*}

\begin{figure*}
\mbox{\includegraphics[width=0.49\textwidth]{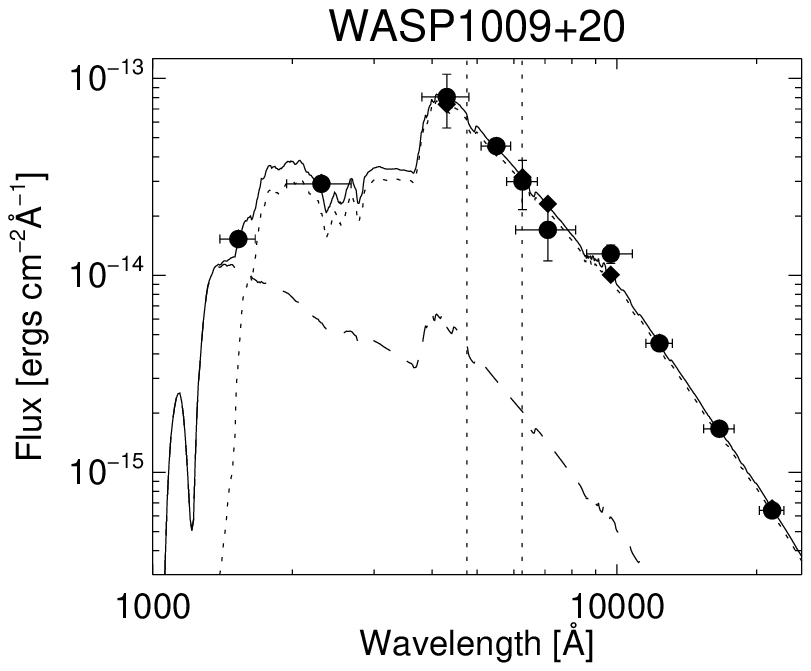}}
\mbox{\includegraphics[width=0.49\textwidth]{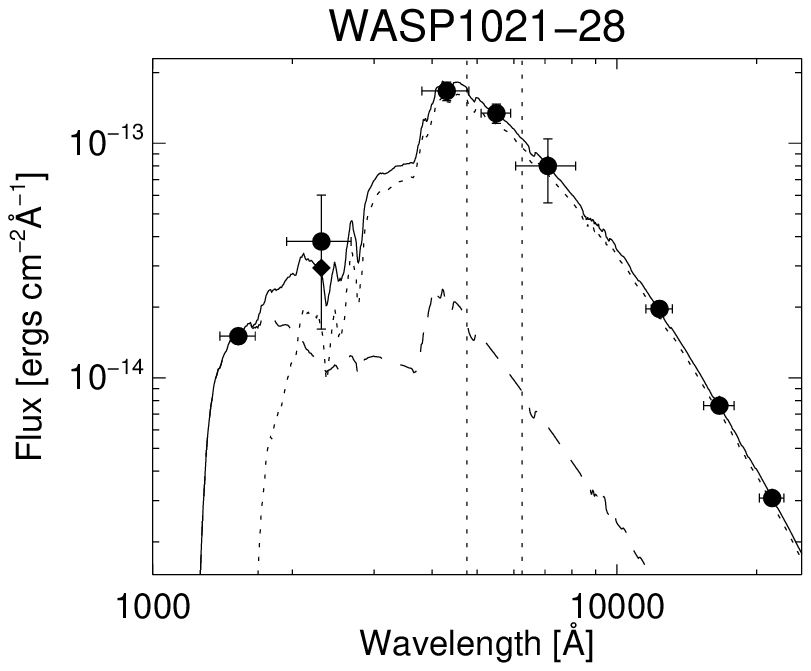}}
\mbox{\includegraphics[width=0.49\textwidth]{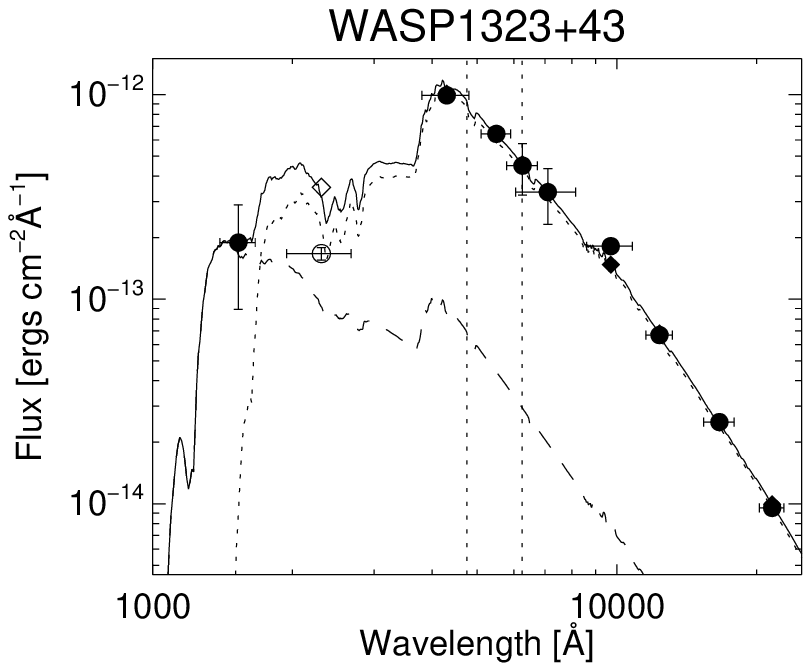}}
\mbox{\includegraphics[width=0.49\textwidth]{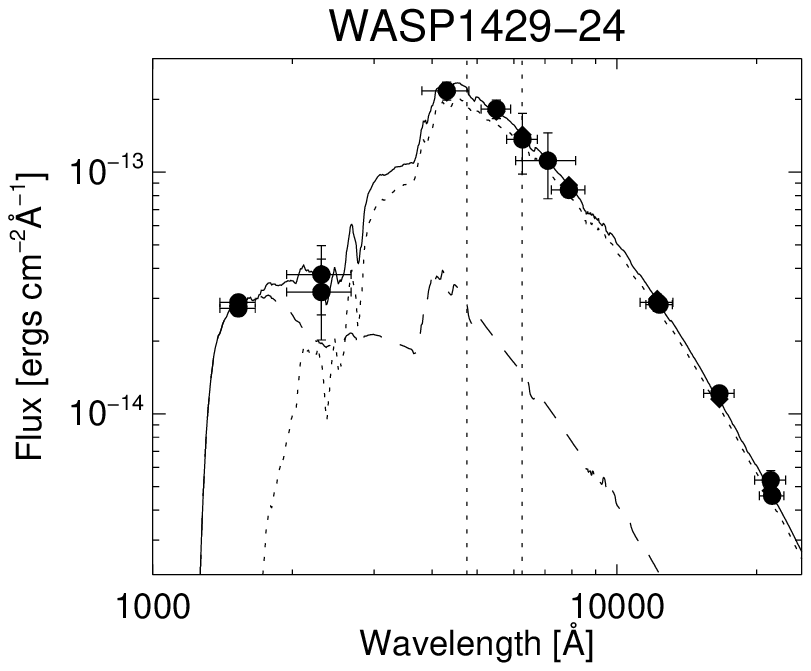}}
\mbox{\includegraphics[width=0.49\textwidth]{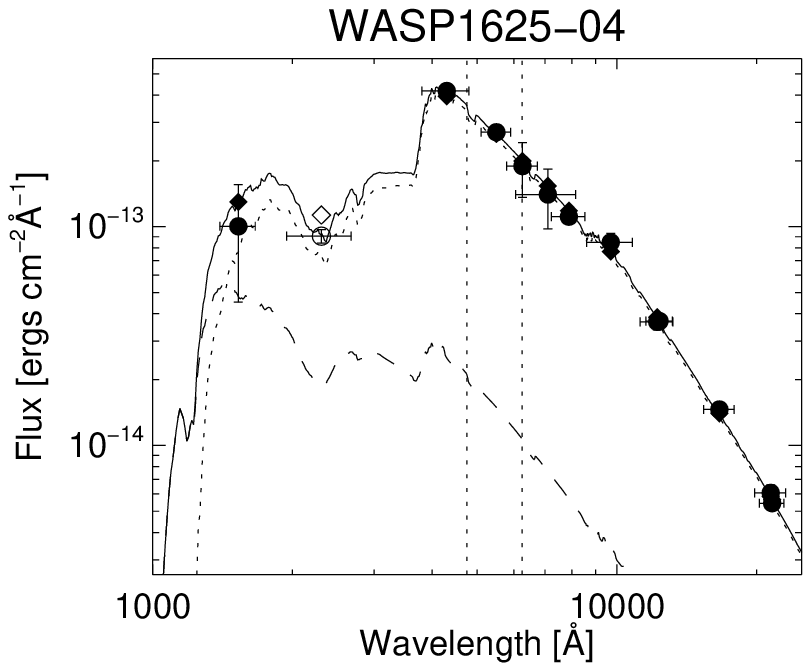}}
\mbox{\includegraphics[width=0.49\textwidth]{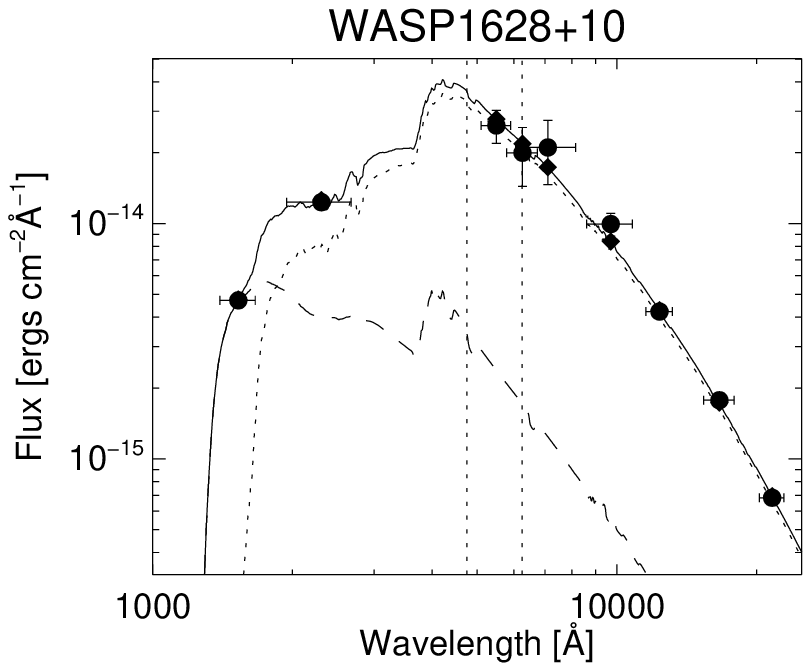}}
\contcaption{}
\end{figure*}

\begin{figure*}
\mbox{\includegraphics[width=0.49\textwidth]{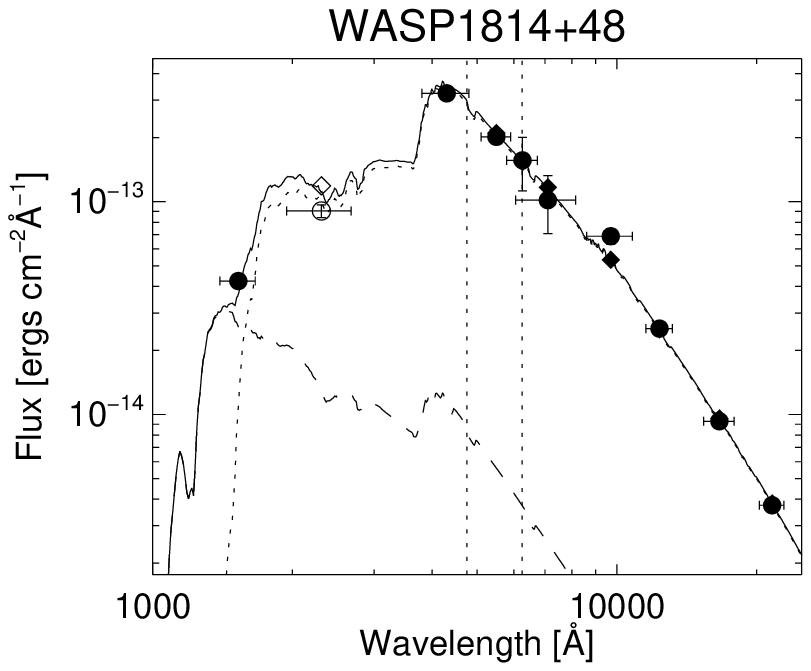}}
\mbox{\includegraphics[width=0.49\textwidth]{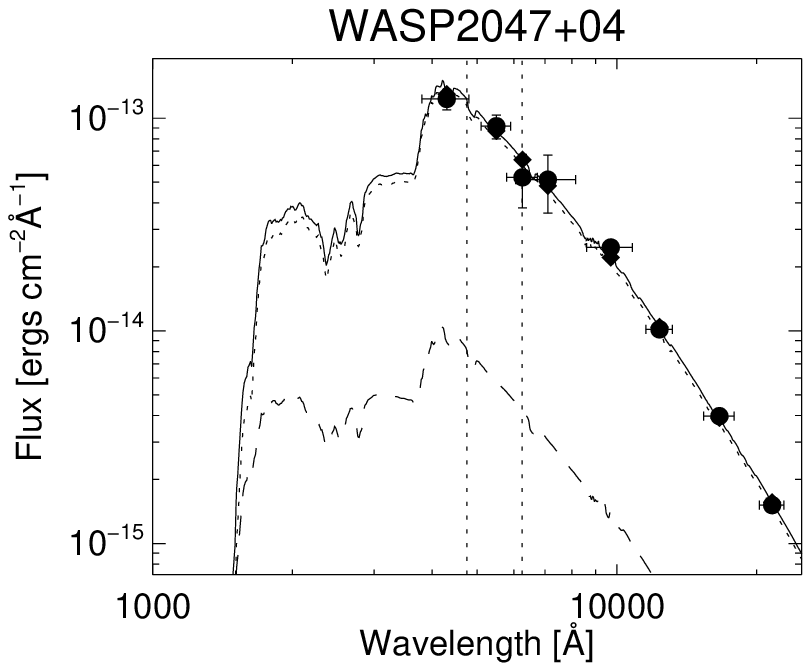}}
\mbox{\includegraphics[width=0.49\textwidth]{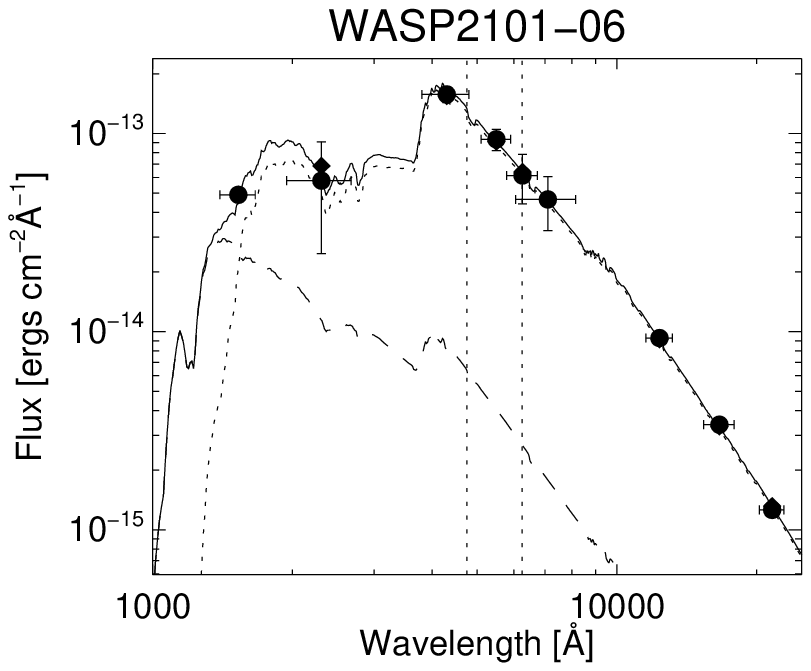}}
\mbox{\includegraphics[width=0.49\textwidth]{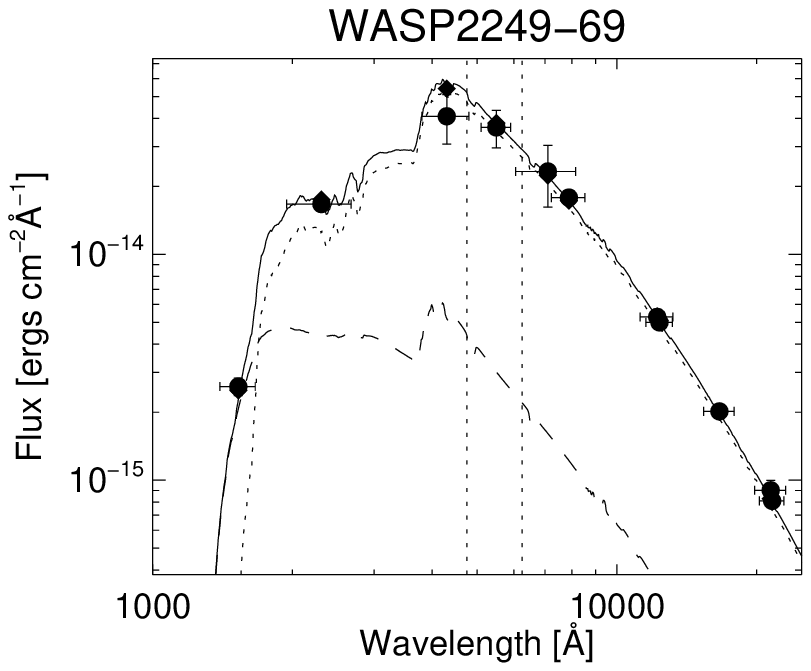}}
\mbox{\includegraphics[width=0.49\textwidth]{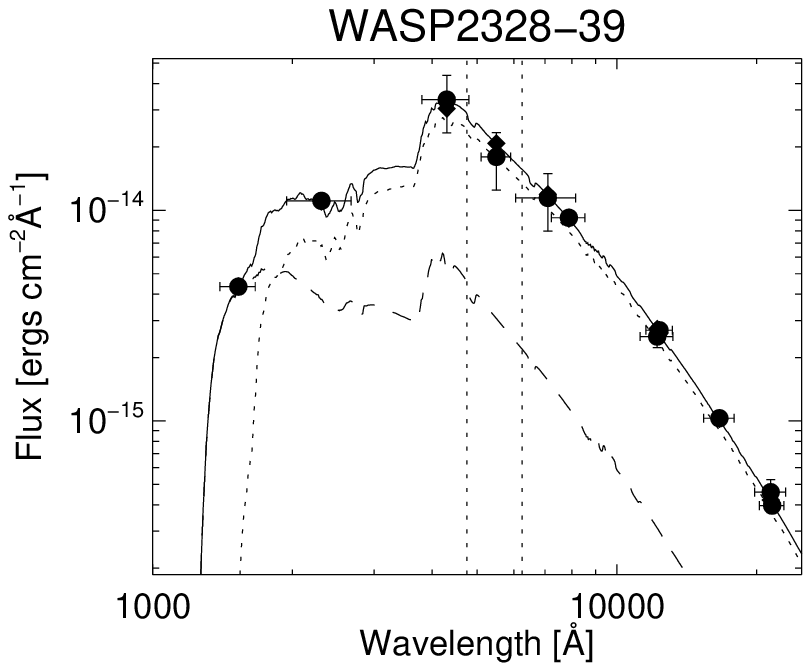}}
\contcaption{}
\end{figure*}
\label{lastpage}

\end{document}